\documentclass{aa}  

\usepackage{graphicx}
\usepackage[dvipsnames]{xcolor}
\usepackage{txfonts}
\usepackage{siunitx} 
\usepackage{natbib,twoopt}
\usepackage[breaklinks=true]{hyperref} %% to avoid \citeads line fills
\bibpunct{(}{)}{;}{a}{}{,}             %% natbib format for A&A and ApJ
\makeatletter
  \newcommandtwoopt{\citeads}[3][][]{\href{http://adsabs.harvard.edu/abs/#3}%
    {\def\hyper@linkstart##1##2{}%
     \let\hyper@linkend\@empty\citealp[#1][#2]{#3}}}
  \newcommandtwoopt{\citepads}[3][][]{\href{http://adsabs.harvard.edu/abs/#3}%
    {\def\hyper@linkstart##1##2{}%
     \let\hyper@linkend\@empty\citep[#1][#2]{#3}}}
  \newcommandtwoopt{\citetads}[3][][]{\href{http://adsabs.harvard.edu/abs/#3}%
    {\def\hyper@linkstart##1##2{}%
     \let\hyper@linkend\@empty\citet[#1][#2]{#3}}}
  \newcommandtwoopt{\citeyearads}[3][][]%
    {\href{http://adsabs.harvard.edu/abs/#3}
    {\def\hyper@linkstart##1##2{}%
     \let\hyper@linkend\@empty\citeyear[#1][#2]{#3}}}
\makeatother
\begin{document}

   \title{Kinematics of metallicity populations in Omega Centauri using {\em Gaia} Focused Product Release and Hubble Space Telescope}
   \authorrunning{N. Vernekar et al.}
 \titlerunning{ Metallicity populations in Omega Centuari}
 
   \author{N. Vernekar
          \inst{1,2}
          \and
          S. Lucatello\inst{2}
          \and
          P. Kuzma\inst{3,4}
          \and
          L. Spina\inst{2}
          }

   \institute{Dipartimento di Fisica e Astronomia, Universitá di Padova, vicolo dell’Osservatorio 2, 35122 Padova, Italy\\
              \email{nagaraj.vernekar@inaf.it}
              \email{nagarajbadarinarayan.vernekar@studenti.unipd.it}
         \and
             INAF - Ossevatorio Astronomico di Padova, vicolo dell’Osservatorio 5, 35122 Padova, Italy
        \and
             National Astronomical Observatory of Japan, 2-21-1 Osawa, Mitaka, Tokyo, 181-8588, Japan
        \and   
            Institute for Astronomy, University of Edinburgh, Royal Observatory, Blackford Hill, Edinburgh, EH9 3HJ, UK}

   \date{Received xxxx; accepted xxxx}

\newcommand{\gaia}{{\em Gaia }}
% \abstract{}{}{}{}{} 
% 5 {} token are mandatory
 
  \abstract
   {Omega Cen is the largest known globular cluster in the Milky Way. It is also a quite complex object with a large metallicity spread and multiple stellar populations. Despite a number of studies over the past several decades, the series of events that led to the formation of this cluster is still poorly understood.
   One of its peculiarities is the presence of a metal-rich population which does not show the phenomenon of light-element anti-correlations (C-N, Na-O, Mg-Al), a trait that is considered as characteristic of Galactic Globular Clusters, and is in fact present among more metal-poor Omega Cen stars, leading to speculations that such anomalous population was accreted by the cluster. 
   %Although its large metallicity spread is unique among Globular Clusters, and moreover it is also due to the presence of the anomalous metal-rich population, which is quite different from the rest of the cluster members, one of the proposed formation mechanisms is through accretion. In this case, a detailed kinematical study of the cluster stars can provide valuable information necessary to obtain constraints on the formation scenario.}
   }
   {In this paper, we aim at investigating the kinematics of Omega Cen populations to gain insight on the formation scenario of the cluster.}
   {Using the newly released \gaia FPR and DR3 catalogue, we conducted a detailed kinematical analysis of cluster members within Omega Cen. The cluster members were divided into four metallicity populations and their mean proper motion in radial and tangential components were compared with each other. We also performed Gaussian Mixture Model fitting on the metallicity distribution to estimate the number of populations within our sample and an independent analysis of the HST catalogue as confirmation.}
   {The mean proper motions ($\mu_{r}$ and $\mu_{t}$) of the metallicity populations do not show any significant differences. It is also not dependent on the approach chosen to determine the number of metallicity populations. We do find clear signature of rotation in all of the populations (including the metal-rich) with similar velocities. }
   {}

   \keywords{Techniques: photometric, globular clusters: individual: Omega Centurari,  Galaxy: kinematics and dynamics 
               }

   \maketitle
%
%-------------------------------------------------------------------
\section{Introduction}
%Past and ongoing spectroscopic surveys such as APOGEE, RAVE, and GALAH combined with exquisite astrometry from Gaia have made it possible to investigate the series of processes that brought to the formation of our Galaxy, leading to a leap ahead in our understanding in the last few years. Upcoming surveys, such as 4MOST, WEAVE and PFS will put us in position to deepen our insight.
%have and will allow us to obtain a more in-depth understanding of our Milky Way Galaxy. In recent years, multiple studies have used the data from these surveys to understand the formation of our galaxy and show the importance of processes such as mergers in shaping our galaxy. 

During a galactic merger, the larger galaxy usually disrupts the dwarf galaxy through tidal interactions \citep{Helmi2001,Mayer2002} and a large percentage of the stars from the dwarf galaxy get dispersed into the Halo of the larger galaxy (in our case, the larger galaxy being the Milky Way). Even with these strong disruption events, the dense central nuclear star clusters (NSCs) of the dwarf galaxies \citep{Nadine2020} can survive and live within the halo of the host galaxy \citep[one example given in ][]{2013MNRAS.433.1997P}. After stripping the external galactic components, the NSC will look similar to a large Globular cluster (GC) \citep{2014MNRAS.441.3570G,Nadine2020} as they have similar mass and radius. The Milky Way has undergone a series of mergers, the last major one being the Gaia-Enceladus satellite merger about 10 Gyr ago \citep{Haywood2018,Helmi2018}. It is also currently undergoing a merger with the Sagittarius dwarf galaxy \citep{Ibata1997,laporte2018}. \cite{pfeffer2014,kruij2019} looked at models of galactic formation and predicted that about 6 ($\pm 1$) clusters in the halo of the Milky Way were likely to be NSC of other galaxies that were ingested. One example that has been identified to be such a stripped NSC is M54, the remnant of Sagittarius dwarf spheroidal galaxy with the stripped stars forming a stream that wraps around the Milky Way \citep{alfaroI,alfaroII,kacharovIII}. 

Omega Centauri (Omega Cen) is quite a unique object in the Milky Way, not only for being the most massive GC in the Milky Way but also due to the complexity of the stellar populations within it. Omega Cen, just like any other GC, was thought to be a simple ensamble of stars formed from the same molecular cloud, with identical ages and formed with the same chemical composition. But in the early 1970s, the evidence of metallicity spread was reported by \cite{1975ApJ...201L..71F}. By the late 1990s, it was also reported to host multiple stellar populations \cite{Anderson1997,Lee1999, Panchino2000, Bedin2004} and more extensive CMD work using the HST \footnote{Hubble Space Telescope}data by \cite{bellini2017a} revealed the presence of multiple stellar populations in all the evolutionary stages. %The combination of large metallicity spread (about 1.6 dex \citep{nitschai2023}) and 
Due to these characteristics, Omega Cen has been suggested be the former nucleus of a dwarf galaxy which was ingested by the Milky Way. Evidence in support of this hypothesis has been put forward in recent years, from the presence of internal stellar disk \citep{VandeVen2006} to possible association with Gaia-Enceladus orbit \citep{Massari2019,pfeffer2021,callingham2022,Limberg2022}, presence of counter-rotating population \citep{Panchetti2024} and an intermediate black hole in the center of the cluster \citep{Haberle2024}. 

High- and intermediate-resolution spectroscopy has been extensively used to characterize the composition of the stellar populations within Omega Cen \citep{Norris1996,sollima2005b,villanova2007,johnson2010,villanova2014,alvarez2024,nitschai2024}. They found that the metal-poor and metal-intermediate populations show anticorrelation in Na-O and Mg-Al abundances, a behaviour consistent with what is observed in all Galactic GCs. 
%and is indicative of the presence of multiple stellar populations within each of these metallicity populations. 
However, the metal-rich population, at $[$Fe/H$]\sim-$0.9\,dex, does not show such anti-correlations, but instead shares the behaviour insofar Na, O, Mg and Al, with field stars. %is the only one that shows a Na-O correlation instead of an anticorrelation. 
This is not only an abnormal behaviour for GC stars, but also completely different from the dominant population of Omega Cen. Due to this, metal-rich has usually been referred to as the anomalous population. 

The origin of the anomalous population is still being debated with at least three scenarios being put forward. 
The first scenario is, ingestion of the field stars of the dwarf galaxy into the central NSC during the disruption event of the merger. This scenario has been linked to the formation mechanism of M54 \citep{caretta2010} which is analogous to Omega Cen. Second scenario is the merger of two cluster (one large metal-poor cluster and another one relatively smaller and metal-rich) and the third scenario is self-enrichment by the first generation Asymptotic Giant Branch (AGB) stars and Type-II supernovae \citep{Norris1996,Smith2000}. Except for the self-enrichment, the other two scenarios require the metal-rich population to be brought into the cluster and therefore, in those cases, the kinematics of the metal-rich stars could still bear the signature of their different origin. In fact, given Omega Cen long relaxation time \citep[about $\sim$10$^{9}$ yr in the center and $\sim$10$^{10}$ yr at half-light radius ][]{1995A&A...303..761M,VandeVen2006}, such signature would be detectable long after the merging/ingestion of the anomalous population, especially when probing the outskirts of the cluster. 
%be different from the stars of the metal-poor cluster, given the large relaxation time of 10 Gyr in the outskirts of Omega Cen.

Previously, \cite{Ferraro2002} used photometry from \cite{Pancino2000} and proper motions from \cite{Leeuwen2000} to investigate the proper motions of the four metallicity populations and concluded that the mean motion of the metal-rich population was not consistent with the other populations and therefore, it might have been the accreted one. But the result was contested by \cite{Platais2003}, who claimed the difference in proper motions reported by \cite{Leeuwen2000} was caused by instrumental effects, whereas \cite{Hugues2004} argued the results were valid because \cite{Leeuwen2000} had made necessary correction. Later, \cite{bellini2009b} used ground-based data and \cite{bellini2018,libralato2018} both used HST data and found no significant difference in the motions of the populations. Even though HST data was considerably more precise than the ground-based data, it included only a small region in the exterior of the cluster. \cite{Sanna2020} conduced the same analysis of the motions but using the \gaia DR2 data for the RGB stars and again found no significant differences. 
The availability of Gaia DR3 and FPR data combined with the recent publication of the 
MUSE data based oMEGACat catalogues \citep{nitschai2023,omegacatII} allows to revisit the issue.

In this study, we use photometry and metallicities from \cite{nitschai2023} and astrometry from \gaia to look at the motions of populations. This gives three advantages over the analysis conducted by previous studies: 1. Complete coverage of Omega Cen, from the outer regions to the dense central core region (see Section \ref{datasets}), 2. Larger sample compared to previous studies and 3. Inclusion of fainter main-sequence stars (upto 2 mag fainter than the turn-off). These advantages allows us to obtained better constraints on the motions of the populations than ever before. In Section \ref{data}, we describe the datasets used for the analysis along with their respective selection criteria, quality cuts and method of cross-matching different catalogues. In Section \ref{analysis}, we provide the procedure of selection of metallicity populations, conversion of proper motions from \gaia to radial and tangential components  and confirmation of the results using another dataset from HST. The interpretation of all the results is given in Section \ref{discussion} with concluding remarks in Section \ref{conclusion}.

%--------------------------------------------------------------------
\section{Data} \label{data}
\subsection{Datasets} \label{datasets}
In order to study the astrometry of stars within Omega Cen, we used the newly released {\em Gaia} Focused Product Release (FPR) \citep{GAIAFPR} (referred to as FPR) catalogue in addition to {\em Gaia} DR3 \citep{GAIADR3} (referred to as DR3). {\em Gaia} FPR was released on 10$^{th}$ October 2023 and in most cases, builds upon the DR3 catalogue by providing astrometry and photometry to many additional sources. 

\begin{figure}
   \centering
   \includegraphics[width=1 \linewidth]{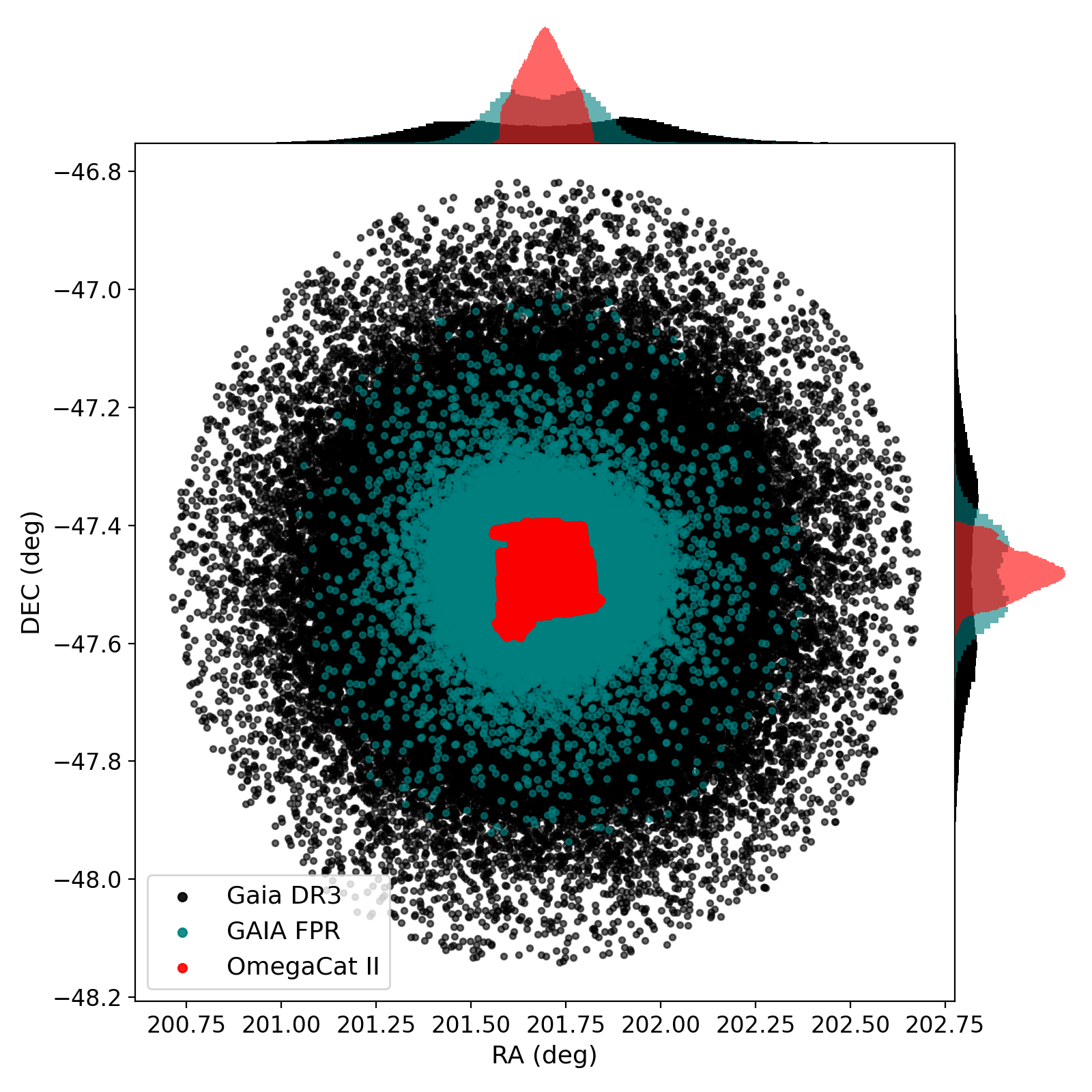}
   \caption{Distribution of stars observed by {\em Gaia} DR3 and \gaia FPR and oMEGACat II catalogue from \cite{omegacatII}. \gaia FPR data is relatively more complete in the central region of Omega Cen in comparison to DR3, whereas oMEGACat II is complete in the very center of the cluster.  }
   \label{fig:densitydistribution}
\end{figure}

In the data released so far by \gaia, there is a limited coverage within dense fields. This is due to the astrometric crowding limitations of the readout window strategy, which is generally used during {\em Gaia}'s nominal mission. This method works well up to a density of 1\,050\,000 objects deg$^{2}$ \citep{GAIAFPR}, but the the density in central regions of GCs such as Omega Cen is significantly higher than this limit. The incompleteness of {\em Gaia} DR3 is shown in Fig. \ref{fig:densitydistribution}. To address this incompleteness issue within the catalogues, the  Gaia Data Processing and Analysis Consortium (DPAC) applied a new software pipeline for the FPR catalogue, where two-dimensional Service Interface Function (SIF) images are used to obtain information about nearby sources and accurately measure the astrometry of faint and crowded stars \citep{GAIAFPR}. {\em Gaia} FPR has a total of 526\,587 objects for the Omega Cen region, while using a cone search of radius 1 deg for Omega Cen within {\em Gaia} DR3, we obtained about 400\,000 stars. 

\begin{figure}
   \centering
   \includegraphics[width=1 \linewidth]{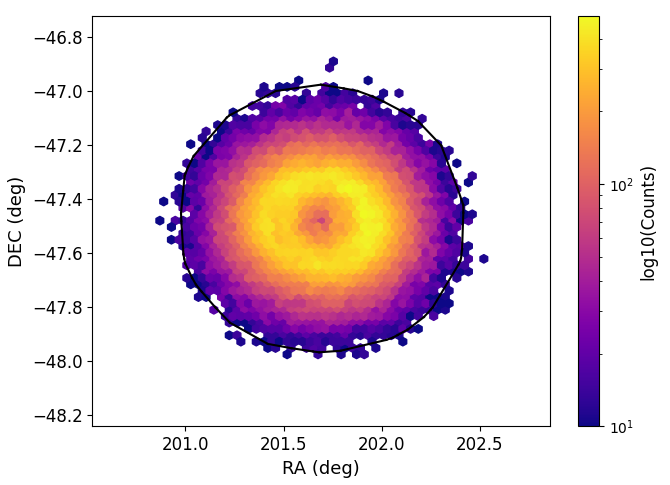}
   \caption{Distribution of \gaia EDR3 stars from \cite{membership} with membership probability $\ge$ 95 $\%$. The colour represents the number of stars within each hexbin. Black circle is the spacial region used to filter \gaia FPR stars.  }
   \label{fig:convexhull}
\end{figure}

\begin{figure}
   \centering
   \includegraphics[width=1 \linewidth]{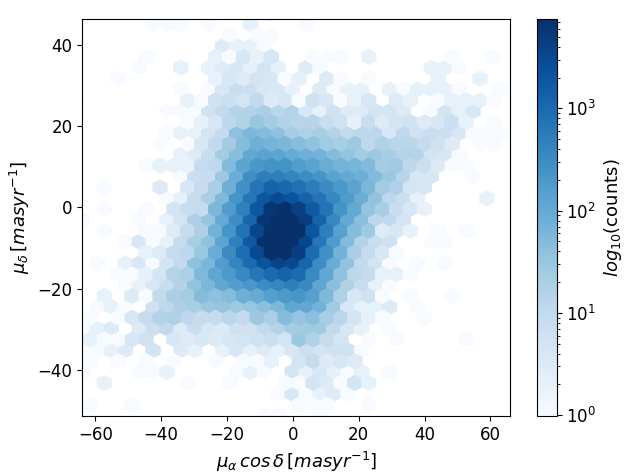}
   \caption{Distribution of proper motions in RA and DEC of all the \gaia FPR stars. Many stars have significantly larger proper motions than the mean proper motion of the cluster. The colour represents the number of stars within each hexbin.}
   \label{fig:fpr_PM}
\end{figure}

The following step was to filter out the non-cluster members from the sample. For the DR3 objects, we adopted the membership probabilities from \cite{membership}, which is based on \gaia EDR3 and set a threshold membership probability of 95 $\%$ or higher for the analysis. As all the FPR stars are new sources with no overlap with the previous {\em Gaia} catalogues, no membership probabilities for them in the literature. Thus, we defined a spatial region using DBSCAN clustering algorithm \citep{dbscan} and convex hull that encompassed the DR3 bona-fide members as shown in Fig. \ref{fig:convexhull} and  
%Therefore, we used stars from \cite{membership} to define a spatial region that primarily encompasses stars with a membership probability greater than 95~$\%$ as shown in Fig. \ref{fig:convexhull}. 
%We 
then considered any star from the FPR catalogue falling within this defined region as a cluster member. In addition to this, we also defined bounds in the proper motions of the stars, where we included stars with proper motions in RA and DEC to be within $\pm$ 5 mas of the cluster's mean proper motions in RA and DEC (taken from \cite{DR2PM}), respectively. The necessity for this selection criterion is due to the presence of stars with highly discrepant proper motions compared to the known motion of the cluster, as seen in Fig. \ref{fig:fpr_PM}. Even though the mentioned method may not seem robust enough, given the high density of cluster members in the interior of the cluster, the percentage of background or foreground stars making it into the final sample should be minimal and may not have a significant effect on the analysis. More importantly, with reliance on the spacial selection criteria and not just on the proper motions, we ensure our final sample will be free of any biases, wherein one of the populations that has a different mean proper motion in comparison to the dominant population of the cluster, could be mislabelled as discrepant and not belonging to the cluster. As for the potential bias caused by the spatial distribution of populations, where centrally concentrated stars from one population are included while those from another population more prevalent in the outskirts are excluded, this is not a concern. \cite{nitschai2024} have shown that the cluster is well-mixed within the half-light radius, thus eliminating any spatial bias. Consequently, our final sample comprises 520,000 stars, with 370,000 from FPR and 150,000 from DR3.

Along with the \gaia catalog, we also analyzed the oMEGACat II catalogue \citep{omegacatII}, which provides precise HST astrometry for about 1.5 million stars in the inner region of Omega Cen. In comparison with \gaia FPR, oMEGACat II not only provides precise astrometry for significantly more centrally located stars (as seen in Fig. \ref{fig:densitydistribution}), it also includes relatively fainter main-sequence stars (upto 26 mag in F435W filter).  

The metallicities and radial velocities (RVs) of the stars were obtained from the oMEGACat I catalogue \citep{nitschai2023}. They used spectroscopic data from MUSE to measure the metallicities and RV of more than 300\,000 stars within the half-light radius of Omega Cen. In addition to this, the catalogue also provides brightness of the objects in two HST Advanced Camera for Surveys (ACS) filters (F435W and F625W).

\subsection{Cross-matching}
%In order to separate the \gaia FPR + DR3 sample into sub-populations, we first needed to cross-match the sample with the oMEGACat I catalogue. 
oMEGACat I is a catalogue based on HST photometry, and it does not include \gaia ID's for the targets. Given that the cluster is a very dense environment, simple position cross-matching between oMEGACAt I and \gaia DR3 and FPR might potentially lead to misidentifications. To minimise these events, we also filtered by the stellar magnitude. As there are no common filters between the photometric systems, we derived transformation equations between the the G magnitude and F435W and F625W passbands.

For this, the \gaia BP/RP (XP) continuous spectra for about 3000 stars from three clusters (M67, NGC\,188 and M15) were used. The choice of these clusters was to cover a wide metallicity range (M67: [Fe/H] $\simeq$ 0.0 dex, NGC\,188: [Fe/H] $\simeq$ 0.1 dex and M15: [Fe/H] $\simeq$ -2.3 dex) and to include a large number of stars in the evolutionary phases. These XP spectra were inputted into the Generator routine of GaiaXPy package to calculate the synthetic HST photometry, including the F435W and F625W band magnitudes. With the two pseudo magnitudes, a simple polynomial curve fitting using eqn. \ref{eqn:calibration} provides the transformation equation. In eqn. \ref{eqn:calibration}, y is (G - F435W) and x is (F625W - F435W) and the values of the coefficients are given in Table \ref{tab:coeff}.

\begin{equation}
     y = Ax^{3} + Bx^{2} + Cx + D
     \label{eqn:calibration}
\end{equation}

\begin{table*}
    \centering
    \caption{Values of coefficients obtained by performing polynomial curve fitting}
    \begin{tabular}{ccccc}
         \hline
         & A & B & C &D\\
         \hline
      G - F435W  & 0.01328 $\pm$ 0.0057  & -0.1114  $\pm$ 0.0292 & 0.6131 $\pm$ 0.0469 & -0.0835 $\pm$ 0.0241\\
      \hline 
    \end{tabular}
    \label{tab:coeff}
\end{table*}

While there is no expectation that this simple approach leads to highly accurate transformation such as in \cite{2023A&A...674A..33G}, it serves the purposes of improving the accuracy of the cross-matching in a dense field.

The cross-matching was performed by adopting a search radius of 0.5\,"  and a magnitude difference of less than 1 mag between the \gaia G-band magnitude and pseudo G-band magnitude of the oMEGACat I catalogue. The low value of the search radius  is due to the high density of stars in the region and a 1 mag photometric threshold was used by taking into account the uncertainties involved in the conversion of HST photometry into pseudo G-band magnitude. With the above criteria, 224\,361 of the 520\,000 objects had a match. We note that the number of matches is not sensitive to the selection criteria used. By doubling the search radius to 1\,", we only obtain about 13,000 (about 6\,\%) additional matches and by increasing the photometric threshold to 1.5 mag, we obtain 18,000 (about 9\,\%) additional matches. This low sensitivity to selection criteria is due the inclusion of both coordinate and photometric search. A colour-magnitude diagram (CMD) of all the cross-matched stars is shown in left panel of Fig. \ref{fig:cmd_selected}.

\subsection{Quality cuts}
\begin{figure*}
   \centering
   \includegraphics[width=1 \linewidth]{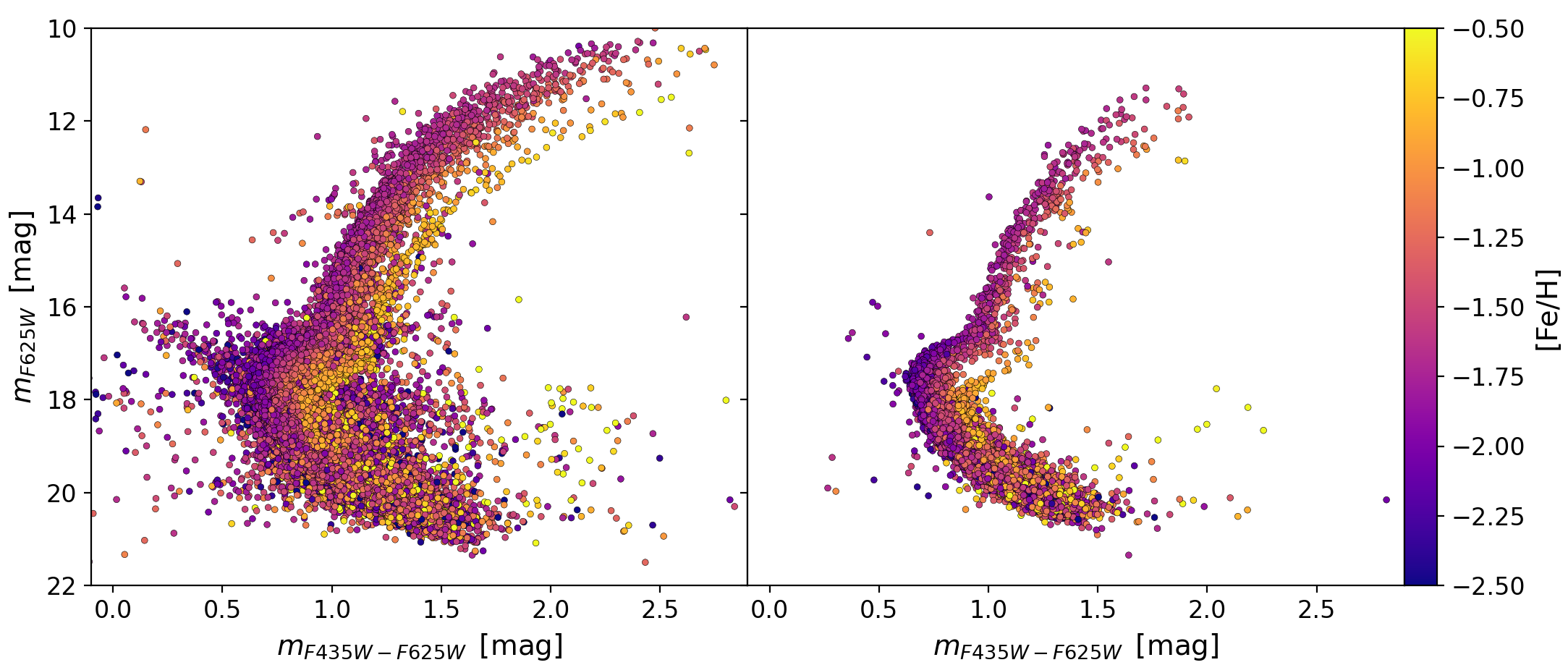}
   \caption{\textit{Left:} CMD of all the stars within our \gaia sample that has a match with the oMEGACat I catalogue. \textit{Right:} CMD of stars that are included in the high-quality sample.}
   \label{fig:cmd_selected}
\end{figure*}

As the aim is to be able to detect even subtle differences in the motions of different populations, it is important to only retain accurate measurements for the analysis. The usual \gaia selection filters were applied, where we only included stars with astrometric excess noise less than 2 and renormalized unit weight error between 0.8 and 1.2. We also refined the sample using the quality flags detailed in \cite{nitschai2023}. These quality cuts brought down our sample significantly, from 224\,361 cross-matched stars to 28,607 stars. This is nearly a 90 \% decrease in the sample size and due to the large astrometric excess noise in majority of the centrally located \gaia FPR stars. These stars most likely suffered from intense crowding which led to large excess noise. The selected stars (referred to as high-quality sample) are plotted on the CMD in the right panel of Fig. \ref{fig:cmd_selected}. 

\section{Analysis} \label{analysis}
\subsection{Metallicity populations}

Literature studies have adopted spectroscopic or photometric approaches to identify various populations in Omega Cen. Interestingly, though, different studies, even when using similar approaches, have reported the existence of varying numbers of metallicity population within the cluster. For example, early study by \cite{1985ApJ...295..437H} reported bimodality in the metallicity distribution in Omega Cen, which was later confirmed by \cite{Norris1996} using calcium abundances and by \cite{1997PhDT.........8A,2004ApJ...605L.125B} using HST photometry. In the early 2000s, studies such as \cite{sollima2005a}, \cite{villanova2007,villanova2014} looked at the sub-giant branch (SGB) of Omega Cen using spectroscopy and reported existence of 4, 4 and 6 populations, respectively. \cite{johnson2010} conduced a spectroscopic study on the red-giant branch (RGB) of Omega Cen and reported 5 populations while \cite{Pancino2000} used photometry and reported 3 populations. \cite{sollima2005b} and \cite{calamida2009} used FORS1 and Stromgren photometry respectively on the RGB to find 5 and 4 populations, respectively. In recent times, \cite{bellini2017} used HST photometry and chromosome maps to photometrically identify at least 15 populations, out of which 9 have been confirmed by \cite{Husser2020}. 
Recently, two studies \cite{alvarez2024} (hereafter AG24) and \cite{nitschai2024} (hereafter SN24) look at the metallicity populations using spectroscopy. AG24 used high-resolution spectroscopic data from VLT/FLAMES to study the Mg-Al anti-correlation with a sample of 439 RGB stars within Omega Cen. They clearly find 4 populations in their metallicity distribution (see Fig. 2 of their paper) with peaks at -1.85, -1.55, -1.15 and -0.80 dex. Whereas, SN24 used metallicities taken from oMEGACat I catalogue \citep{nitschai2023}, which is based low-resolution MUSE data, to study a sample of 11,050 RGB stars, finding 11 different populations. 

Given such a diverse number of populations being reported, we used two different methods to divide our sample into. First method is to use the bounds defined in existing literature, while the second method is to estimate the number of populations directly within our sample.

\subsubsection{Method 1: Literature}
For this analysis, we used the metallicity bounds defined by AG24, the most recent high-resolution study of the chemistry of stars within the cluster. We selected this study over SN24, despite it being the latest research on Omega Cen populations and having the largest sample to date, because SN24's analysis is based on metallicities derived from low-resolution spectroscopic data, resulting in larger measurement uncertainties. In contrast, AG24 utilized spectroscopic data with a resolution of 20,000–29,500 and a signal-to-noise ratio of 70–100. This higher resolution and quality enabled AG24 to better constrain the metallicities of even the most metal-poor stars ([Fe/H] < -2.0 dex), providing more reliable metallicity bounds for the populations.

\begin{figure}
   \centering
   \includegraphics[width=1 \linewidth]{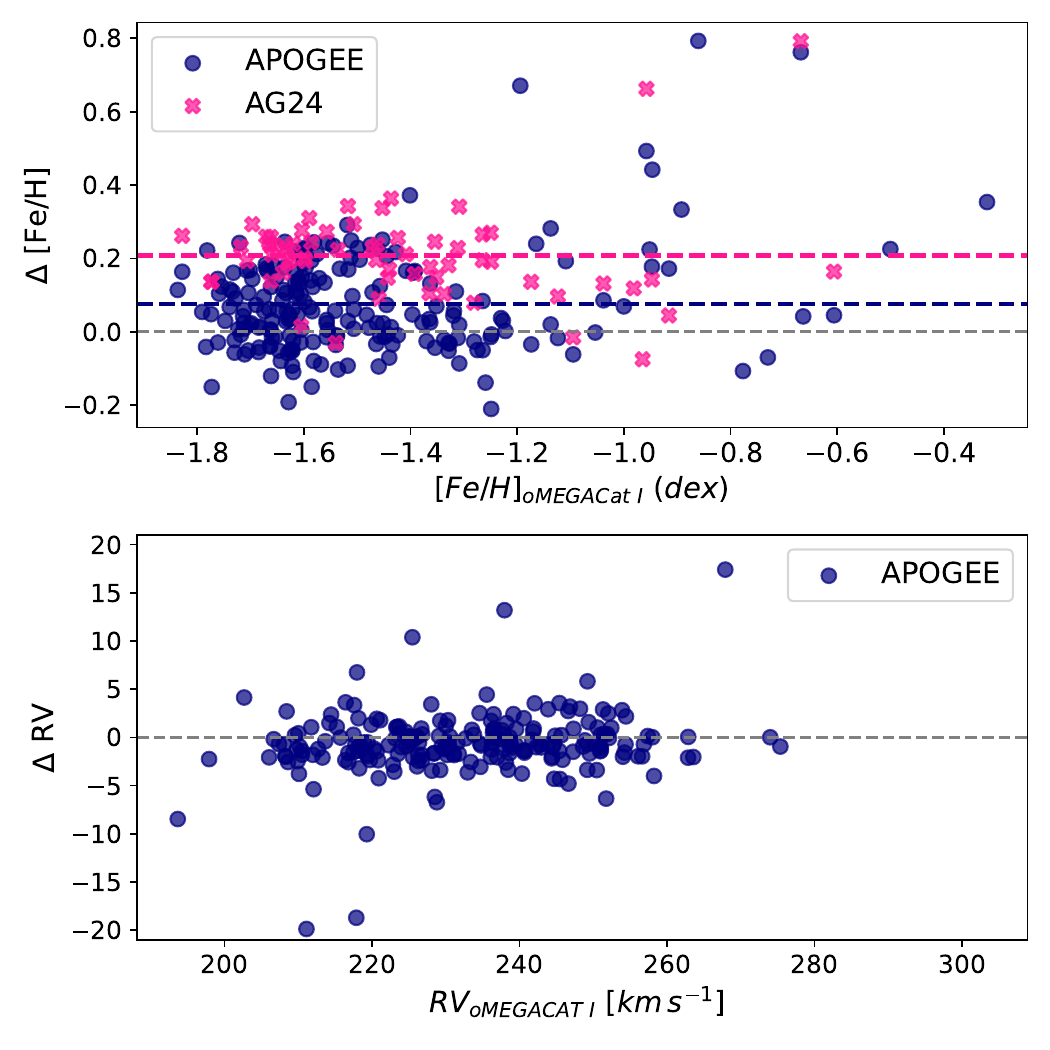}
   \caption{\textit{Top:} Comparison between metallicities of oMEGACat I catalogue with that from AG24 (pink cross) and APOGEE (purple circles). The purple and pink dashed-line represent the mean offset of AG24 and APOGEE from oMEGACat I metallicities, respectively. \textit{Bottom:} Comparison of radial velocities between oMEGACat I and APOGEE. The y axis of the top and bottom panel represent [Fe/H]$_{oMEGACat I}$ - [Fe/H]$_{lit}$ and RV$_{oMEGACat I}$ - RV$_{lit}$, respectively.}
   \label{fig:nitvsalv}
\end{figure}

Before dividing our sample into sub-populations following AG24, we probed the existence of any offsets or trends in the metallicities from oMEGACat I (used for the analysis) with respect to AG24 (used to define bounds). In the top panel of Fig. \ref{fig:nitvsalv}, we compare oMEGACat I metallicities with AG24 and APOGEE\footnote{Apache Point Observatory Galactic Evolution Experiment} \citep{APOGEE2020}. We find a mean metallicity difference between AG24 and oMEGACat I to be about 0.2 dex. Therefore, to account for this offset within the analysis, we adopted a correction factor of 0.2 dex on the oMEGACat I metallicities. With respect to APOGEE, the mean metallicity difference is smaller at 0.07 dex. In the bottom panel, we also compare the RV from oMEGACat I and APOGEE, where we find a good agreement between the two.

\begin{figure*}
   \centering
   \includegraphics[width=1 \linewidth]{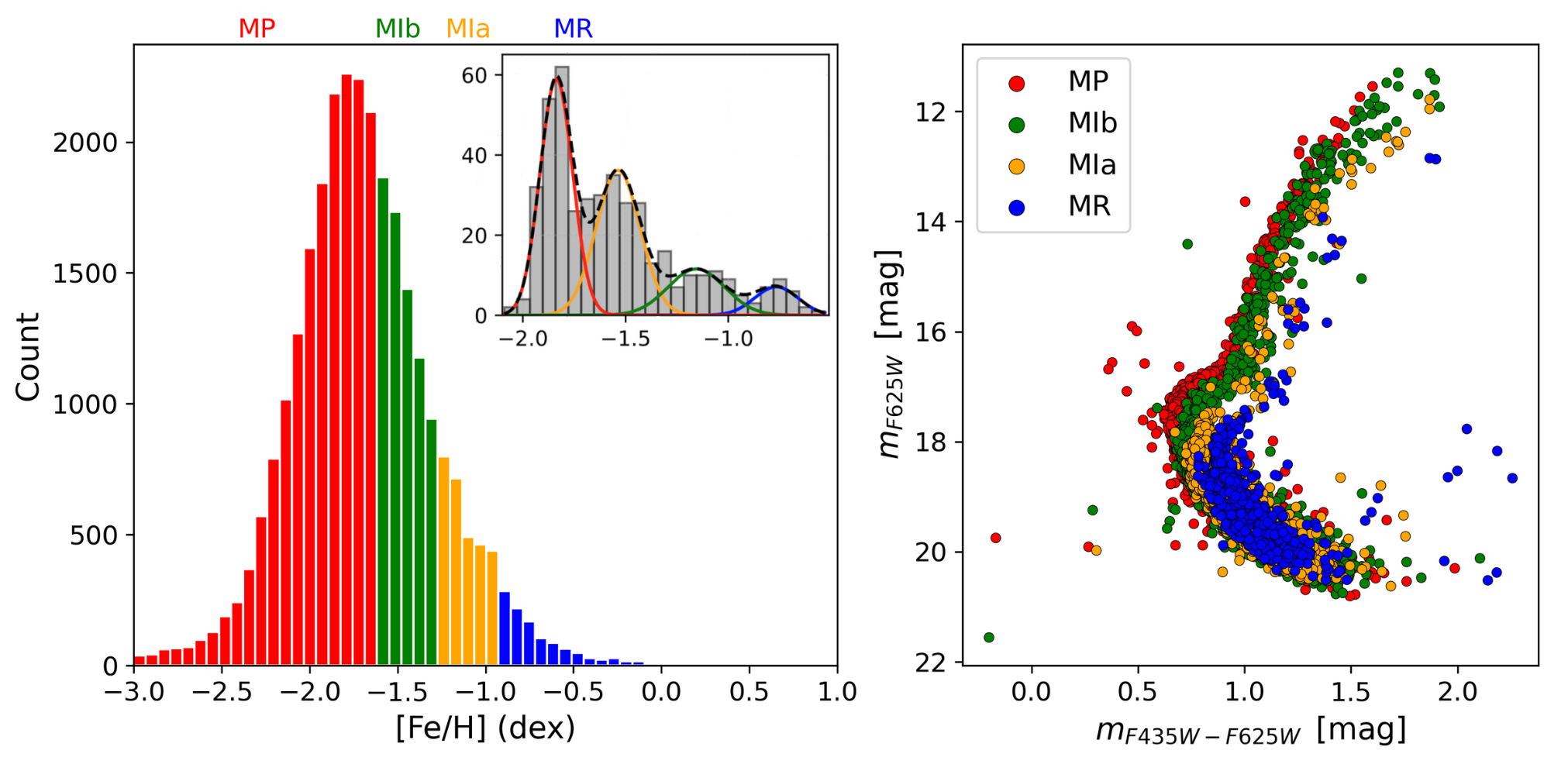}
   \caption{\textit{Left:} Metallicity distribution of our sample with the one from AG24 shown in the inserted panel. Different metallicity populations are represented in different colors. \textit{Right:} CMD of the different sub-populations.}
   \label{fig:alv_mh_cmd}
\end{figure*}

Following AG24, we divide our sample into four sub-populations defined as: metal-rich (MR, [Fe/H] $\ge$ -0.93 dex), metal-intermediate a (MIa, -1.33 $\le$ [Fe/H] $<$ -0.93 dex ), metal-intermediate b (MIb, -1.68 $\le$ [Fe/H] $<$ -1.33 dex ) and metal-poor (MP, [Fe/H] $<$ -1.68 dex). The metallicity distribution of the four populations and their corresponding locations in the CMD are shown in Fig. \ref{fig:alv_mh_cmd}, (left and right panels, respectively). 

The metallicity distribution and the four sub-populations are shown in left panel of Fig. \ref{fig:alv_mh_cmd} with the CMD of the four sub-populations in the right panel of Fig. \ref{fig:alv_mh_cmd}.

From Fig. \ref{fig:alv_mh_cmd}, it can be noticed that the metallicity distribution of our sample is quite different from the one in AG24. This could be due to several factors, such as larger uncertainties on the metallicities in the oMEGACat I catalogue, different sample size, or different stellar types (our sample includes RGB, turn-off stars as well as main sequence stars that are upto 3 mag fainter than turn-off in the F625W filter band, whereas AG24 only includes RGB stars).  

\begin{figure*}[h]
   \centering
   \includegraphics[width=1 \linewidth]{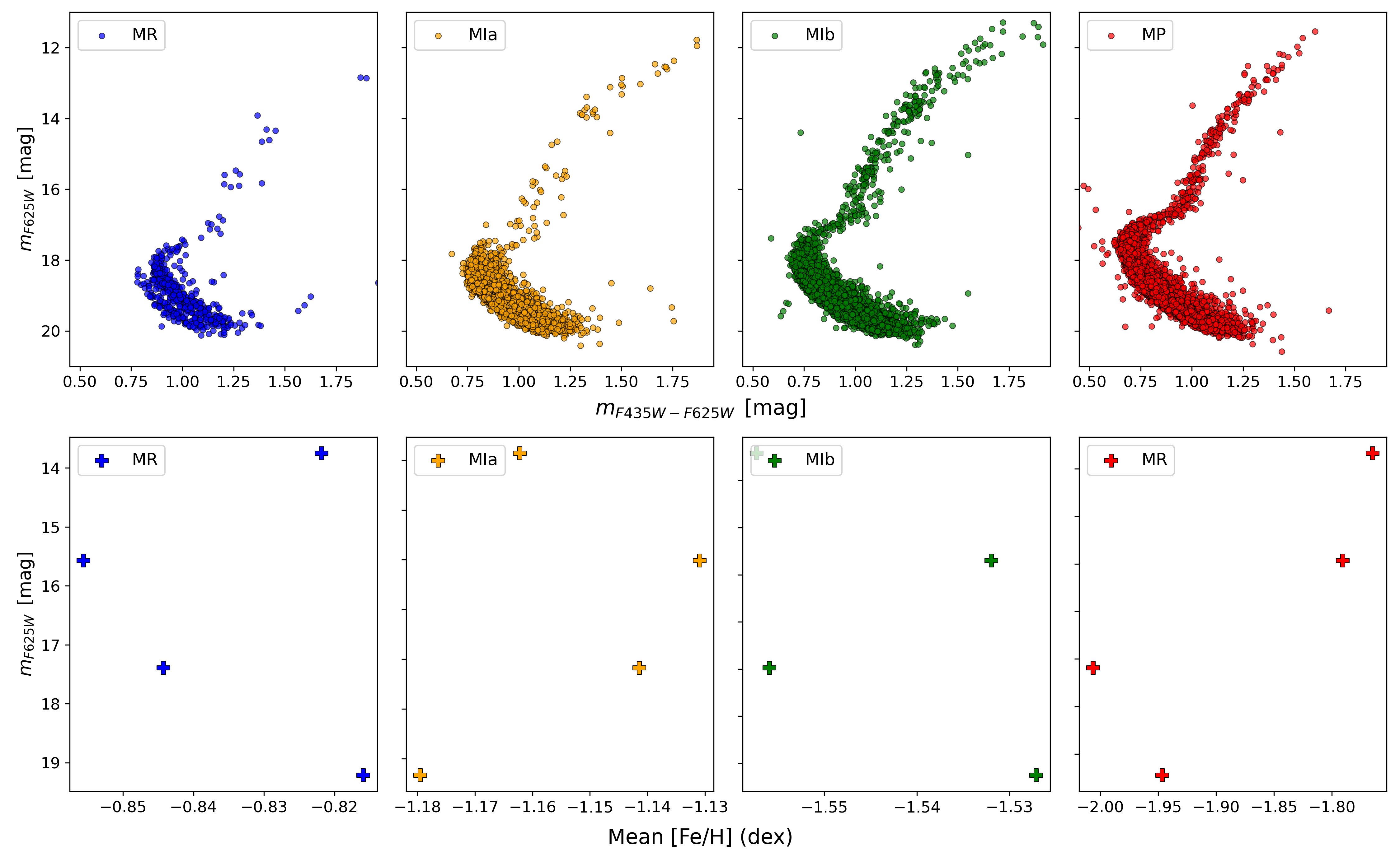}
   \caption{\textit{Top: }Individual CMDs of populations from Metal Rich on the left most panel to Metal Poor on the right most panel. \textit{Bottom: }Mean metallicities of the magnitude bins within different populations.}
   \label{fig:mhoffset}
\end{figure*}

In Fig. \ref{fig:mhoffset}, the top panel shows the individual CMDs of the populations. All of the populations have well defined main sequence, but only the metal intermediate b and metal poor populations have well populated RGB with metal intermediate a and metal rich populations having only a few stars. The bottom panel shows the mean metallicity of different magnitude bins within each populations. In three populations, i.e. MR, MIa and MIb, the maximum difference between metallicities corresponding to MS and RGB is about 0.04, 0.05 and 0.03 dex respectively. But in MP population, this difference is about 0.18 dex (with RGB being more metal rich), with it increasing to 0.24 dex between turn-off and RGB. This offset could be attributed to the difficulty in parameter estimation at such low-metallicities using a low resolution spectrum.

\subsubsection{Method 2: Metallicity distribution fitting}

\begin{figure}[h]
   \centering
   \includegraphics[width=1 \linewidth]{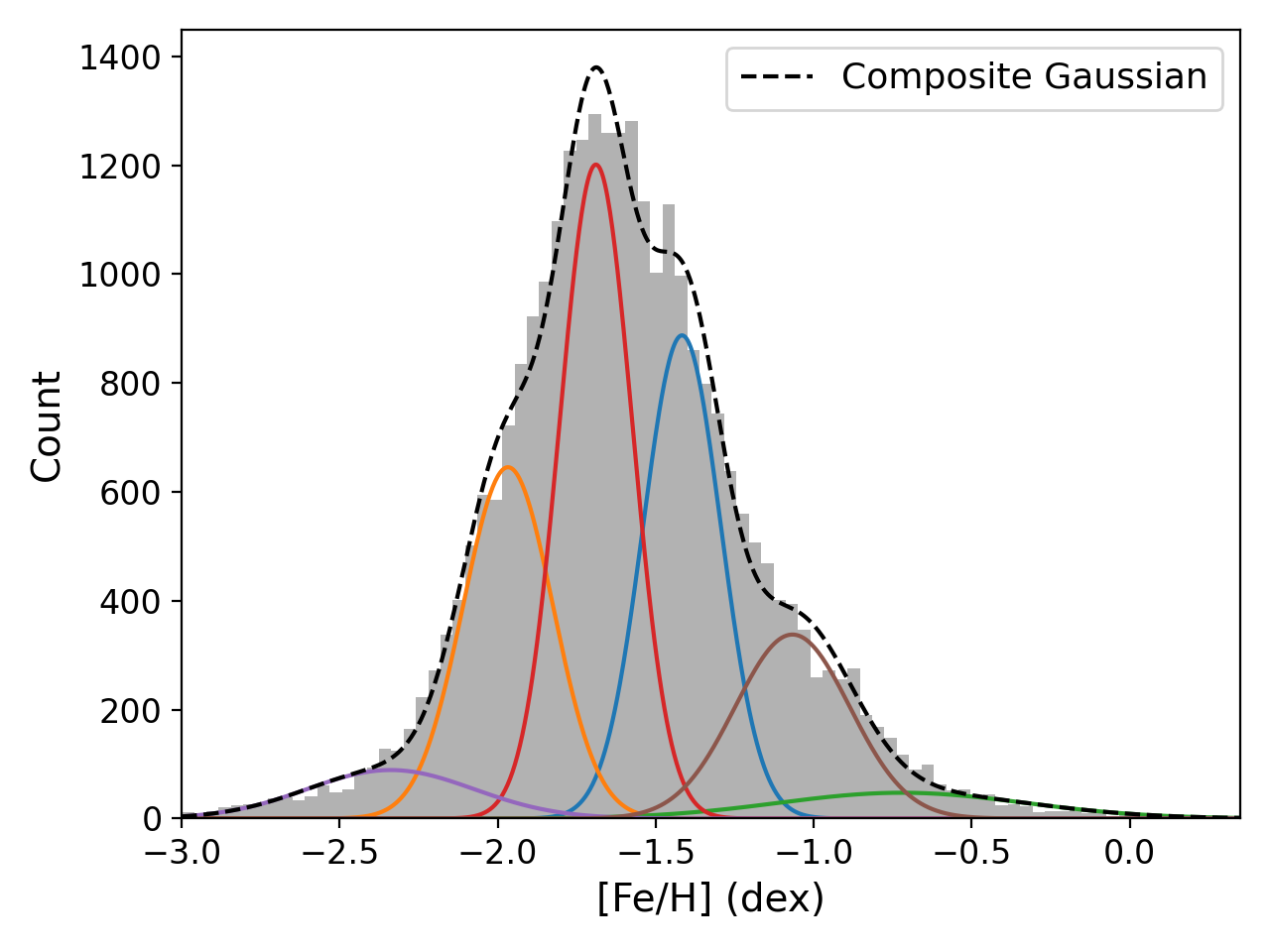}
   \caption{Metallicity distribution of the sample with 6 different Gaussian components plotted using different colours and representing the six sub-populations within Omega Cen. }
   \label{fig:gmm_hist}
\end{figure}

An alternative approach is to estimate the number of sub-populations within our sample by fitting the metallicity distribution using the Gaussian Mixture Model (GMM) provided under the Scikit-learn\footnote{https://scikit-learn.org/stable/modules/mixture.html} python package \citep{scikit-learn}. We fit the metallicity distribution with GMMs consisting of different number of components (ranging from 2 to 20 components). The best fit model is selected using the Bayesian Information Criterion (BIC), which in our case is 6 components. The fitting of the metallicity distribution and individual components of the best fit is shown in Fig. \ref{fig:gmm_hist} with the results of the fitting in Table \ref{tab:GMM_para}. The six peaks are at [Fe/H]: -2.35, -1.95, -1.60, -1.25, -1.05 and -0.60 dex. For comparison, the four peaks found in AG24 are at -1.85, -1.55, -1.15 and -0.80 dex. Three of the six peaks in this study, i.e. -1.95, -1.60 and -1.25 dex are within 1$\sigma$ of the first three peaks of AG24, with our values being larger. The peak at -1.05 dex is slightly more discrepant compared to -0.80 dex of AG24 but is within the 3$\sigma$ range. Along with these four, we also find one most metal-poor peak and another most metal-rich peak at -2.35 and -0.60 dex, respectively. The most metal-poor peak is consistent (within 1$\sigma$) with the metal-poor population found in \cite{johnson2020} at -2.25 dex. But this peak in \cite{johnson2020} only contained 11 stars, whereas in this case it amounts to over 900 stars. However, the two sub-populations at -2.35 and -0.60 dex constitute less than 6 $\%$ of the total sample, with 68 $\%$ of the stars having a [Fe/H] between -1.78 and -1.03 dex.

\begin{table}
    \centering
        \caption{Components obtained by fitting GMM on the metallicity distribution}
    \begin{tabular}{cccc}
    \hline
        Mean [Fe/H]  & Number of & Fraction \\
          (dex)    &   stars &  $\%$\\
          \hline
        -2.35 $\pm$ 0.20  & 908  & 3.19 \\
        -1.95 $\pm$ 0.13 & 5492 & 19.30\\
        -1.60 $\pm$ 0.10 & 9758 & 34.28\\
        -1.25 $\pm$ 0.11 & 7532 & 26.45\\
        -1.05 $\pm$ 0.15 & 4058 & 14.25\\
        -0.60 $\pm$ 0.25 & 721  & 2.53\\
         \hline
    \end{tabular}
    \label{tab:GMM_para}
\end{table}

\subsection{Proper motion}

\begin{figure*}
   \centering
   \includegraphics[width=1 \linewidth]{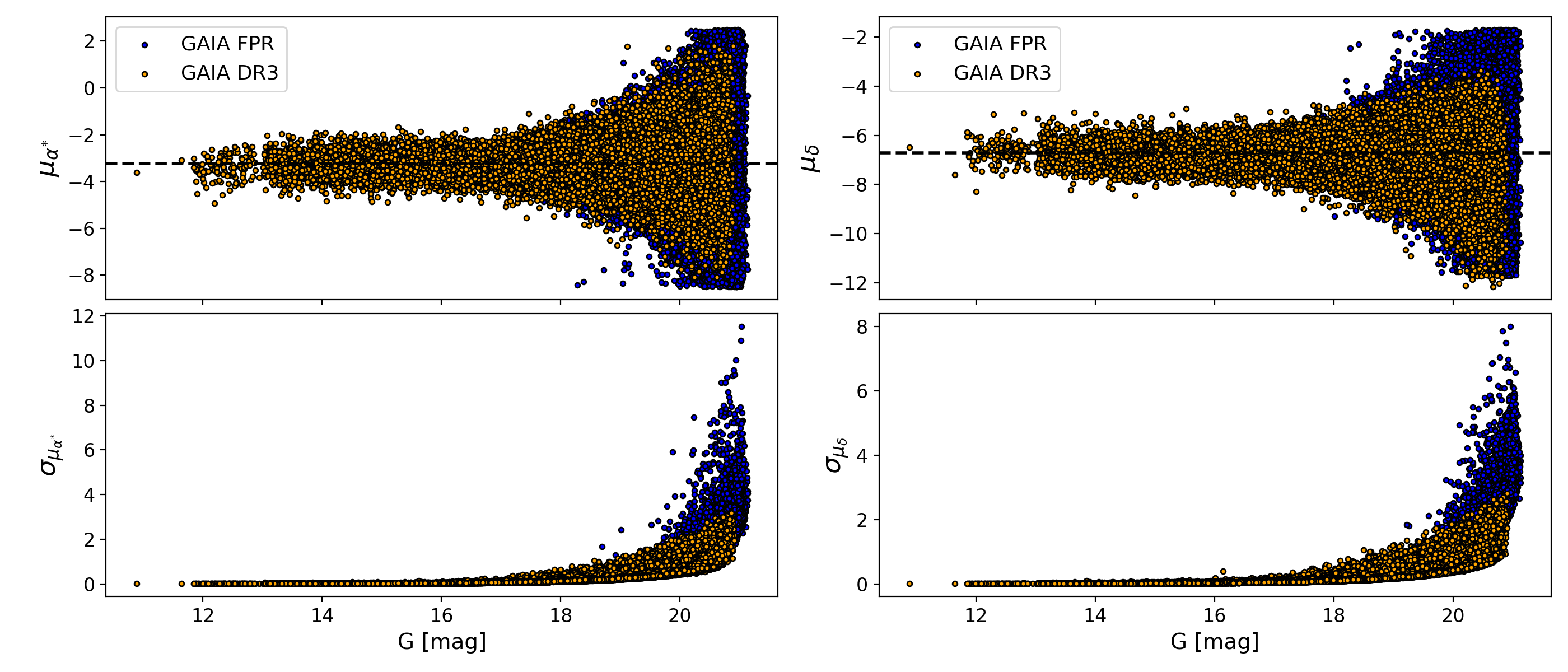}
   \caption{\textit{Top: } Proper motions in RA (left) and DEC (right) as function of the G-band magnitudes. \textit{Bottom: } Errors associated with proper motions in RA (left) and DEC (right) as function of the G-band magnitude of the star. Stars in blue are taken from \gaia FPR catalogue and stars in yellow are taken from \gaia DR3.}
   \label{fig:pm}
\end{figure*}

We used the \gaia proper motions ($\mu^{*}_{\alpha}$\footnote{$\mu^{*}_{\alpha}$ = $\mu_{\alpha}$\,$\cos{\delta}$ } and $\mu_{\delta}$) to analyse the motions of the sub-populations. Given the high sensitivity needed to decipher small difference between the populations, we investigated the precision of the \gaia measurements, specifically the proper motions components. In Fig. \ref{fig:pm}, we plot $\mu^{*}_{\alpha}$ and $\mu_{\delta}$ in the upper panels with their corresponding errors in the lower panels. Based on \gaia DR2 data, the mean proper motion values for Omega Cen is $\mu^{*}_{\alpha}$ = 3.24 $mas\,yr^{-1}$ and $\mu_{\delta}$ = -6.73 $mas\,yr^{-1}$ \citep{DR2PM} and is represented with a black dashed line in Fig. \ref{fig:pm}. It is evident from Fig. \ref{fig:pm}, that the proper motion measurements are more likely to deviate from the average value as we go towards fainter stars. This is an expected behaviour as it is difficult to accurately identify and accurately measure the position of faint stars when it is in a dense field. In addition to this deviation, we also see the uncertainties on the measurements rapidly increase as we go fainter magnitudes. This effect is exaggerated in \gaia FPR compared to DR3, as the former provides data mostly for objects close to the core of the cluster. To remove any adverse effects of these large uncertainties from our analysis, we decided to use only the best 20 $\%$ of the measurements in each magnitude bins. By doing this, we not only remove the highly uncertain measurements, but also use reliable stars in the fainter end of the CMD for the analysis. This is important as no previous study using spectroscopic metallicities has attempted to look at Omega Cen's sub-populations using main sequence stars about 3 magnitudes fainter than turn-off (has been performed by \citep{bellini2018} using photometry).

To evaluate differences in the motions of populations within the cluster, it is advantageous to transform the proper motions from \gaia to radial and tangential components referenced at the center of the cluster. Doing so, will address projection effects caused due to curvature of the sky and rotation in the plane of the sky, while also subtracting the mean motion of the cluster center from individual stellar motions. Thereby such a transformation will result in maximising any differences in the mean motions of the populations with respect to each other. The method of the transformation is as follows:

\begin{itemize}
    \item Transform the equatorial coordinates ($\alpha$ and $\delta$) into cartesian coordinates (X', Y') using eqn. \ref{cartesian} (taken from \cite{VandeVen2006}), where $r_{0}$ = 10800/$\pi$ and ($\alpha_{0}$, $\delta_{0}$) is the equatorial center of Omega Cen with $\alpha_{0}$ = 201.697 deg and $\delta_{0}$ = -47.480 deg \citep{membership}. This transformation is applicable to objects with large angular diameter such as a GC. In this transformation, positive x-axis denotes west direction.

    \begin{equation}
    \begin{aligned}
    X' &= -r_0 \cos(\delta) \sin(\alpha - \alpha_0) \\
    Y' &= r_0 \left( \sin(\delta) \cos(\delta_0) - \cos(\delta) \sin(\delta_0) \cos(\alpha - \alpha_0) \right)
    \end{aligned}
    \label{cartesian}
    \end{equation}
\vspace{0.1cm}

    \item Transform the proper motions ($\mu_{\alpha^{*}}$, $\mu_{\delta}$) into Cartesian coordinate system using eqn. \ref{propermotion}. To make the coordinates consistent with the motions, we need to change $\mu_{x}$ to -$\mu_{x}$ as proper motions are positive in west direction.

    \begin{equation}
    \begin{aligned}
-\mu_{x} &=  \mu_{\alpha} \cos(\alpha - \alpha_{0}) - \mu_{\delta} \sin(\delta) \sin(\alpha - \alpha_{0}) \\
\mu_{y}  &=  \mu_{\alpha} \sin(\delta_{0}) \sin(\alpha - \alpha_{0}) \\
         &\quad + \mu_{\delta} (\cos(\delta) \cos(\delta_{0}) + \sin(\delta) \sin(\delta_{0}) \cos(\alpha - \alpha_{0}))
    \end{aligned}
    \label{propermotion}
    \end{equation}

   \item Before converting the Cartesian projections of proper motions into radial and tangential components, they need to be corrected for perspective rotation using eqn. 6 of \cite{VandeVen2006}. This is due to a combination of large angular diameter of Omega Cen and its systematic motion that results in apparent rotation that has to be corrected for. Even though the influence of perspective rotation is larger at the outskirts of the cluster, we still chose to correct for it within our sample. 

    \item Use the corrected Cartesian projections in eqn. \ref{rad_tang} (taken from \cite{2000A&A...360..472V}) to obtain the proper motions in radial and tangential projections. In eqn. \ref{rad_tang}, R is the projected distance from the cluster center and the uncertainties are propagated with the help of \cite{2019MNRAS.484.2832V}

    \begin{equation}
    \begin{aligned}
        \mu_{r} &= (x \mu_{x} + y \mu_{y}) / R\\
        \mu_{t} &= (-y \mu_{x} + x \mu_{y}) / R
    \end{aligned}
    \label{rad_tang}
    \end{equation}

  % \begin{equation}
   %  \begin{aligned}
   % \centering
    %    \sigma_{\mu_{x}}^2 &=  [\sigma_{\mu_{\alpha}} \cos(\alpha - \alpha_{0})]^2 + [-\sigma_{\mu_{\delta}} \sin(\delta) \sin(\alpha - \alpha_{0})]^2 \\
    %    & - 2 \sigma_{\mu_{\alpha}} \sigma_{\mu_{\delta}}\,\rho\,[\sigma_{\mu_{\alpha}} \cos(\alpha - \alpha_{0})] [\sigma_{\mu_{\delta}} \sin(\delta) \sin(\alpha - \alpha_{0})] \\
    %    \sigma_{\mu_{y}}^2 &=  [\sigma_{\mu_{\alpha}} \sin(\delta_{0} \sin(\alpha - \alpha_{0})]^2 \\
    %    &+ [\sigma_{\mu_{\delta}} \cos(\delta) \cos(\delta_{0}) \sin(\delta) \sin(\delta_{0}) \cos(\alpha - \alpha_{0})]^2 \\ 
    %    &+ 2 \sigma_{\mu_{\alpha}} \sigma_{\mu_{\delta}}\,\rho\,\sigma_{\mu_{\delta}} \cos(\delta) \cos(\delta_{0}) \sin(\delta) \sin(\delta_{0}) \cos(\alpha - \alpha_{0})] \\
%        &[\sigma_{\mu_{\delta}} \cos(\delta) \cos(\delta_{0}) \sin(\delta) \sin(\delta_{0}) \cos(\alpha - \alpha_{0})]
 %   \end{aligned}
  %  \label{x_y_err}
   % \end{equation}

 %   \begin{equation}
%    \begin{aligned}
%    \centering
%        \sigma_{\mu_{r}} &= \sqrt{\frac{(\sigma_{\mu_{x}} \cos(\text{arctan2}[\frac{y}{x}]))^2}{(\sigma_{\mu_{y}} \sin(\text{arctan2}[\frac{y}{x}]))^2}} \\
%        \sigma_{\mu_{t}} &=  \sqrt{\frac{(\sigma_{\mu_{x}} \sin(\text{arctan2}[\frac{y}{x}]))^2}{(\sigma_{\mu_{y}} \cos(\text{arctan2}[\frac{y}{x}]))^2}}
%    \end{aligned}
%    \label{rad_tang_err}
%    \end{equation}

\end{itemize}

\begin{figure}[h]
   \centering
   \includegraphics[width=0.99 \linewidth]{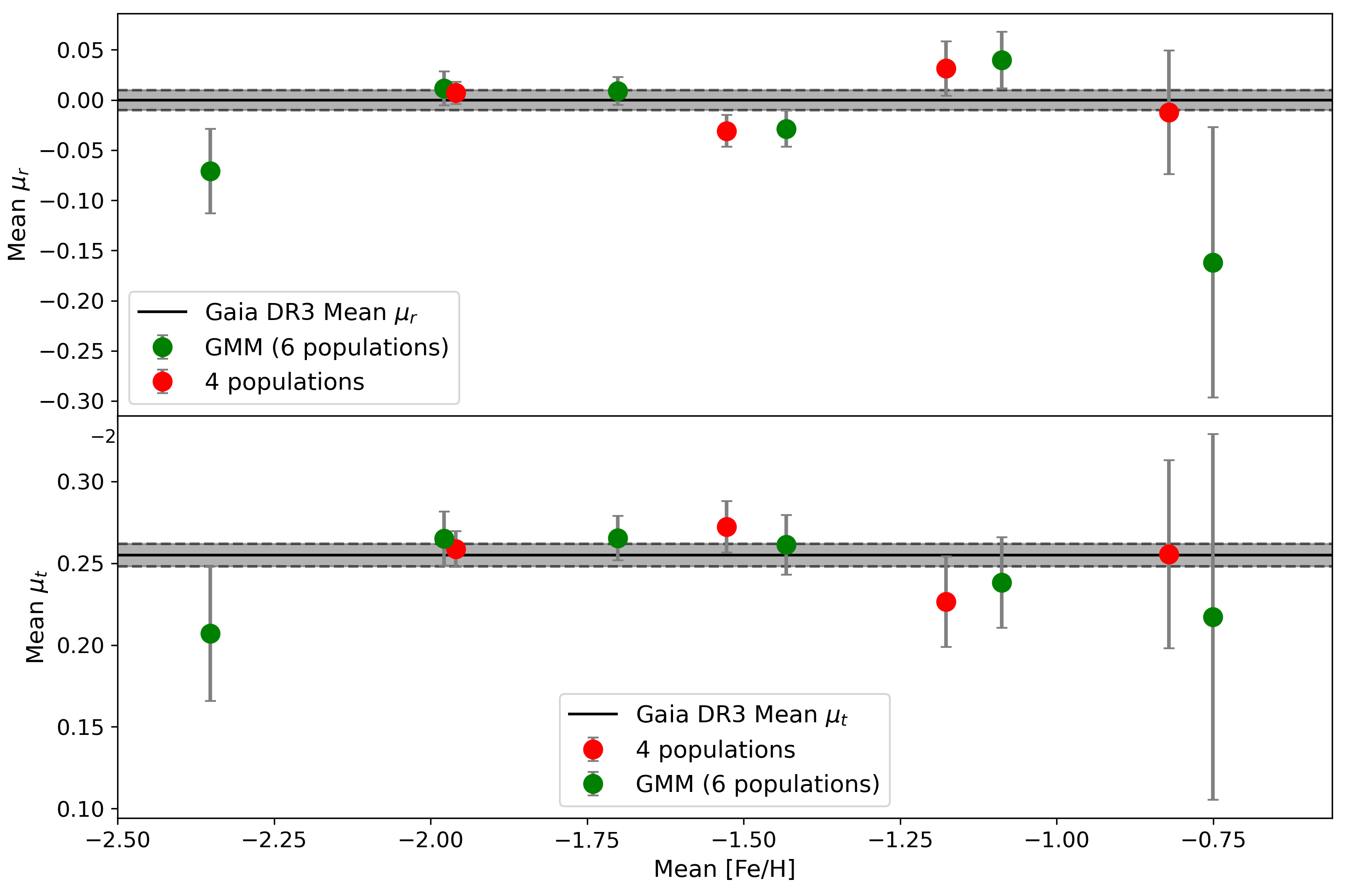}
   \caption{Mean radial (top) and tangential (bottom) proper motions of the sub-populations obtained using AG24 (4 populations, red circles) and GMM (6 populations, green circles). The black line represents the mean values of proper motion calculated using \gaia DR3 stars.}
   \label{fig:4pop}
\end{figure}

\begin{figure*}[h]
   \centering
   \includegraphics[width=1 \linewidth]{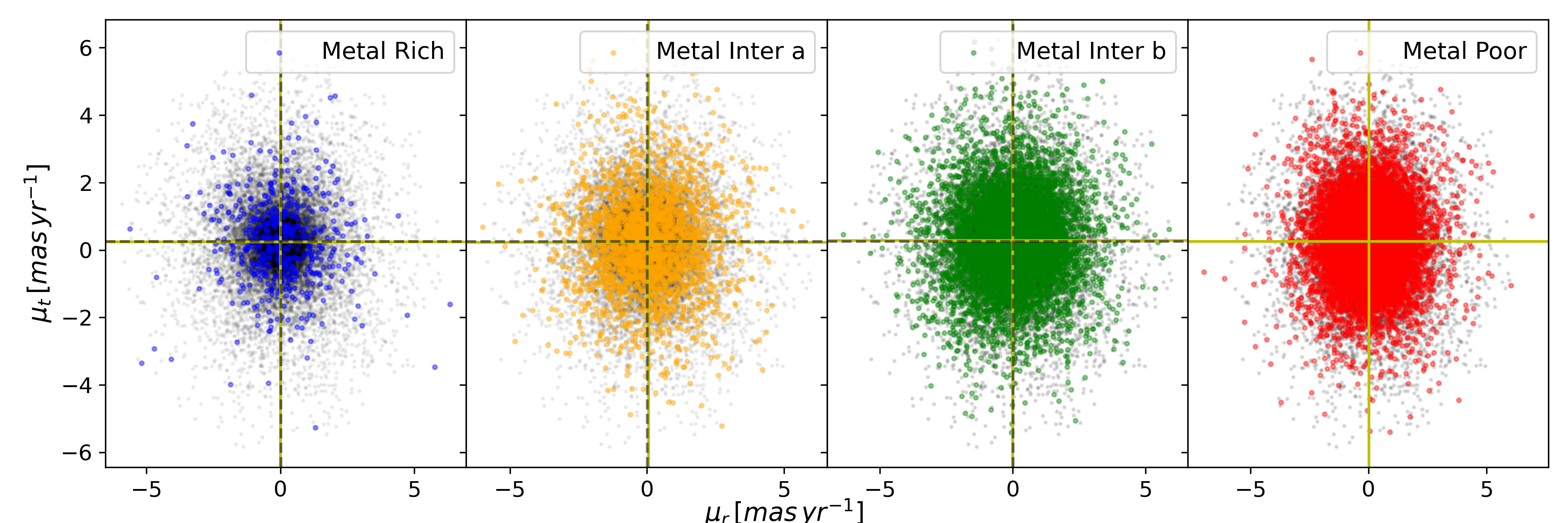}
   \caption{Radial and tangential components of the proper motions of individual stars in the four populations obtained using AG24. The vertical and horizontal yellow line represents the mean values of radial and tangential components of a particular population, respectively. The black dashed vertical and horizontal line represents the mean values of radial and tangential component of the metal-poor population.}
   \label{fig:pmr_pmt}
\end{figure*}

We considered the top 20 $\%$ of the proper motions in radial and tangential components with the smallest uncertainties and calculated the mean motions for the sub-populations. The average proper motions for the four populations obtained using metallicity bounds from AG24 are shown in Fig. \ref{fig:4pop} along with the mean proper motions calculated using the 150\,000 \gaia DR3 stars with membership probability higher than 95 \%.The motions of individual stars in each of the populations are shown in Fig. \ref{fig:pmr_pmt}. The differences between the motions of the populations are poorly significant, with one of the metal intermediate population showing the largest difference, which is still however of limited significance (less than 2$\sigma$).  The value for the MR  population is consistent with the others, even if it has large error associated with it. Similar considerations apply when separating the sample in the six populations obtained from GMM method, we do not see any significant differences in the motions as seen in Fig. \ref{fig:4pop}.

\subsection{Rotation}

Early study by \cite{norris1997} analysed a sample of 400 RGB stars to understand the kinematical differences between the metal rich and metal poor populations. They used radial velocities from \cite{mayor1997} and calcium abundances from \cite{Norris1996} to differentiate the stars into two populations: Metal Poor (80 $\%$ of the total sample, [Ca/H] $\le$ -1.2 dex) and Metal Rich (20 $\%$ of the sample, [Ca/H] $>$ -1.2 dex). They found the metal rich population to be more centrally located as well as to be kinematically cooler than the metal poor population. But most importantly, they report that the metal poor population shows well-defined systemic rotation, whereas the metal rich population is a non-rotating population (see Fig. 3 in their paper). Later, \cite{pancino2007} looked at this rotation problem using high-resolution FLAMES spectra for about 650 RGB stars. They divided their sample into three populations (metal rich, metal intermediate and metal poor). They found all the three populations to show the same rotational patterns. The measurement for the MR population, however, was based on a limited number of stars (only 70 stars or about 11 \% of their sample) and presented a large scatter. Therefore, any signature of rotation that was being reported was associated with a large uncertainty.  

\begin{figure}[h]
   \centering
   \includegraphics[width=1 \linewidth]{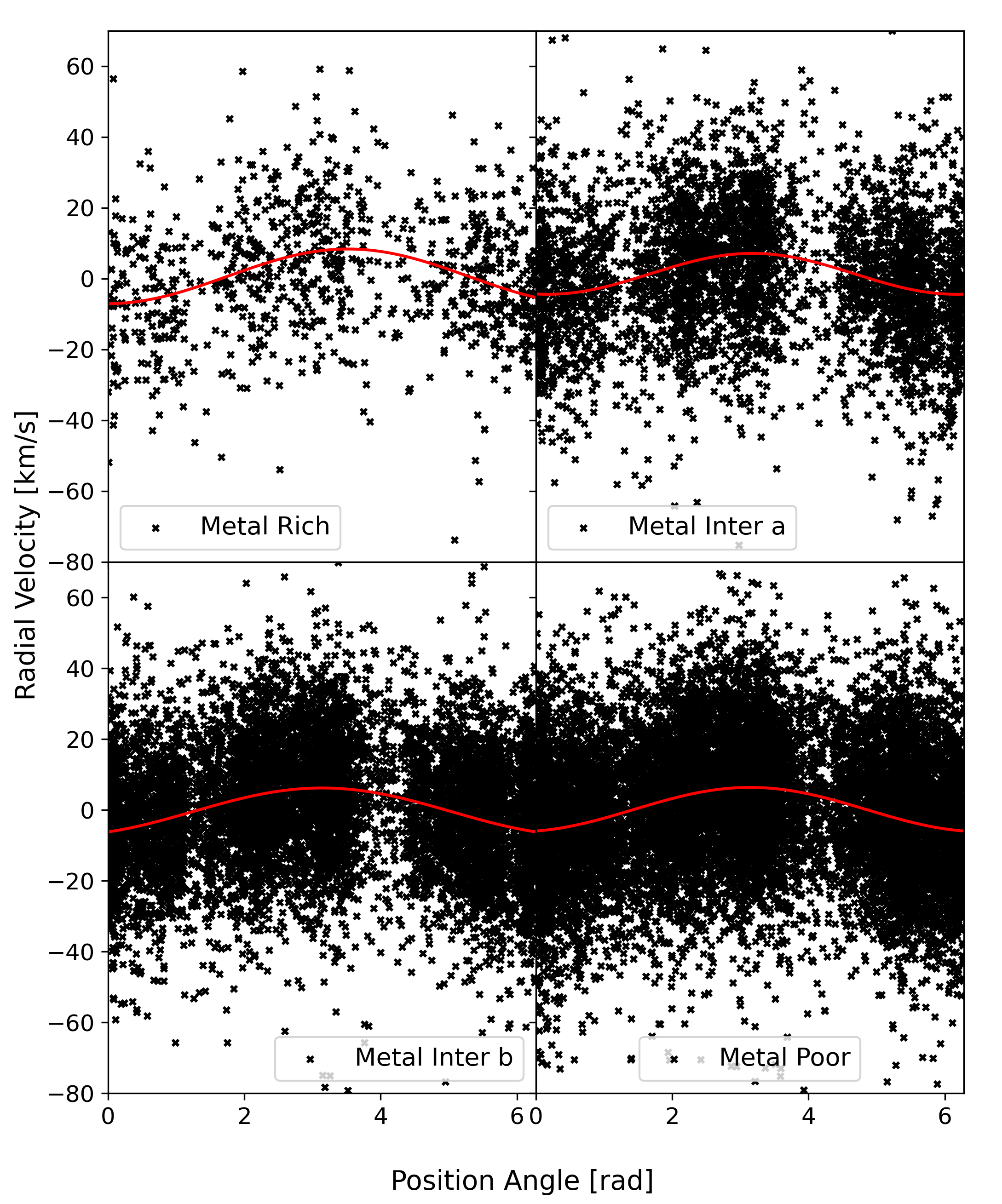}
   \caption{Radial velocity plotted against position angle. The red line represents the best non-linear fit to the distribution. The sinusoidal variation represents the rotational motion within the populations.}
   \label{fig:4pop_RV}
\end{figure}

\begin{figure}
   \centering
   \includegraphics[width=1 \linewidth]{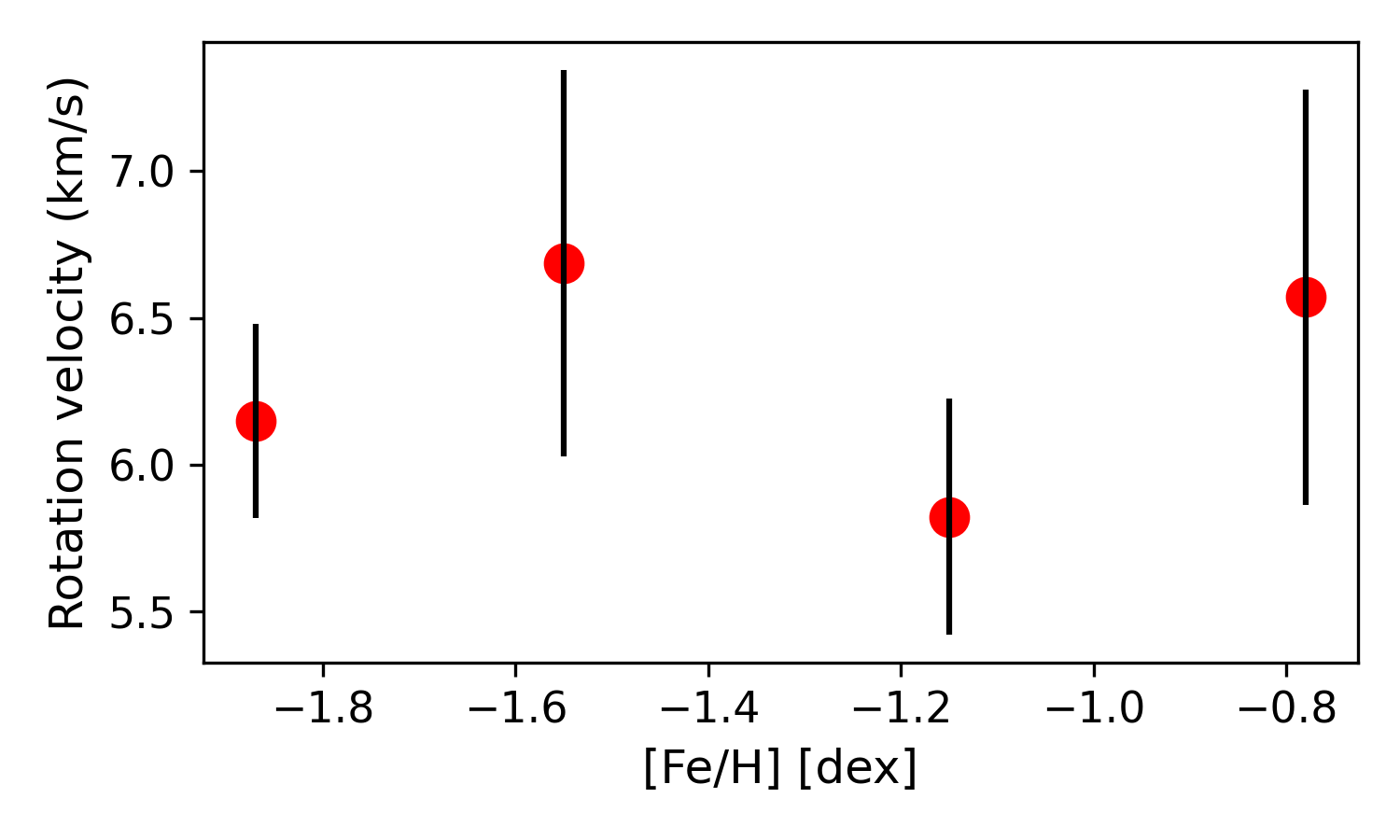}
   \caption{Rotational velocity of the four metallicity populations obtained using AG24. The values plotted above are the amplitude of the non-linear fitting performed in Fig. \ref{fig:4pop_RV}. }
   \label{fig:rotation}
\end{figure}

We revisit the issue taking advantage of our large sample (even for metal rich population) and conduct a similar analysis as \cite{pancino2007}. The first step was to calculate the position angle for all the stars within the sample as this would clearly represent the rotation with a sinusoidal variation. Once the position angles were obtained, we plotted it against the RVs obtained from oMEGACat I catalogue as shown in Fig. \ref{fig:4pop_RV} . We made sure to include only reliable measurements for the RVs where the value was $\ge$ 3$\,\sigma$ and subtract the mean RV of the cluster from the individual measurements. The red curve in Fig. \ref{fig:4pop_RV} is a non-linear fit obtained using $\chi^{2}$ minimisation. We clearly identify the sinusoidal variation in all of the four populations, implying the presence of rotational motion even in the most metal rich population of Omega Cen. From the non-linear fitting, we estimate the amplitudes of the rotation velocities of the populations. Similar to the results from \cite{pancino2007}, we find rotational velocities to be uniform (as per Fig. 
\ref{fig:rotation}) among the populations and the amplitudes to be around 6 km/s. The rotational velocities reported in Fig. \ref{fig:rotation} are only indicative values as we expect stars with different radial distance to have different rotational values. 

\subsection{oMEGACat II}

All the analysis mentioned above makes use of stars within the \gaia sample (FPR and DR3). As we already discussed, crowding is an issue in the central region of the cluster, due to which the \gaia sample had relatively low number of stars with accurate astrometry in the center. 
To better probe the center of the cluster
%we  Therefore, to make sure the results we found in the above mentioned analysis holds true in the very center of the cluster, 
we conducted the same analysis using HST data obtained from oMEGACat II catalogue. This catalogue provides precise astrometry and multi-band photometry for over 1.5 million stars in the cluster's core (covers a region of (0.2 x 0.2) deg around the center, see Fig. \ref{fig:densitydistribution}). The astrometry in this catalogue is not only significantly better than Gaia FPR in terms of the uncertainties on the proper motions (see Fig. 12 of \cite{omegacatII}), but it also includes stars $\sim$4 mag fainter than the faintest stars within the Gaia sample. The only limitation of this sample is its coverage which is restricted to the very center and does not include any stars from the outer regions of the clusters. Inclusion of some outer regions could allow for a comparison study between the inner and outer region stars in terms of kinematics as \cite{norris1997} had reported the centrally concentrated metal rich population was kinematically cooler. 

\begin{figure}[h]
   \centering
   \includegraphics[width=1 \linewidth]{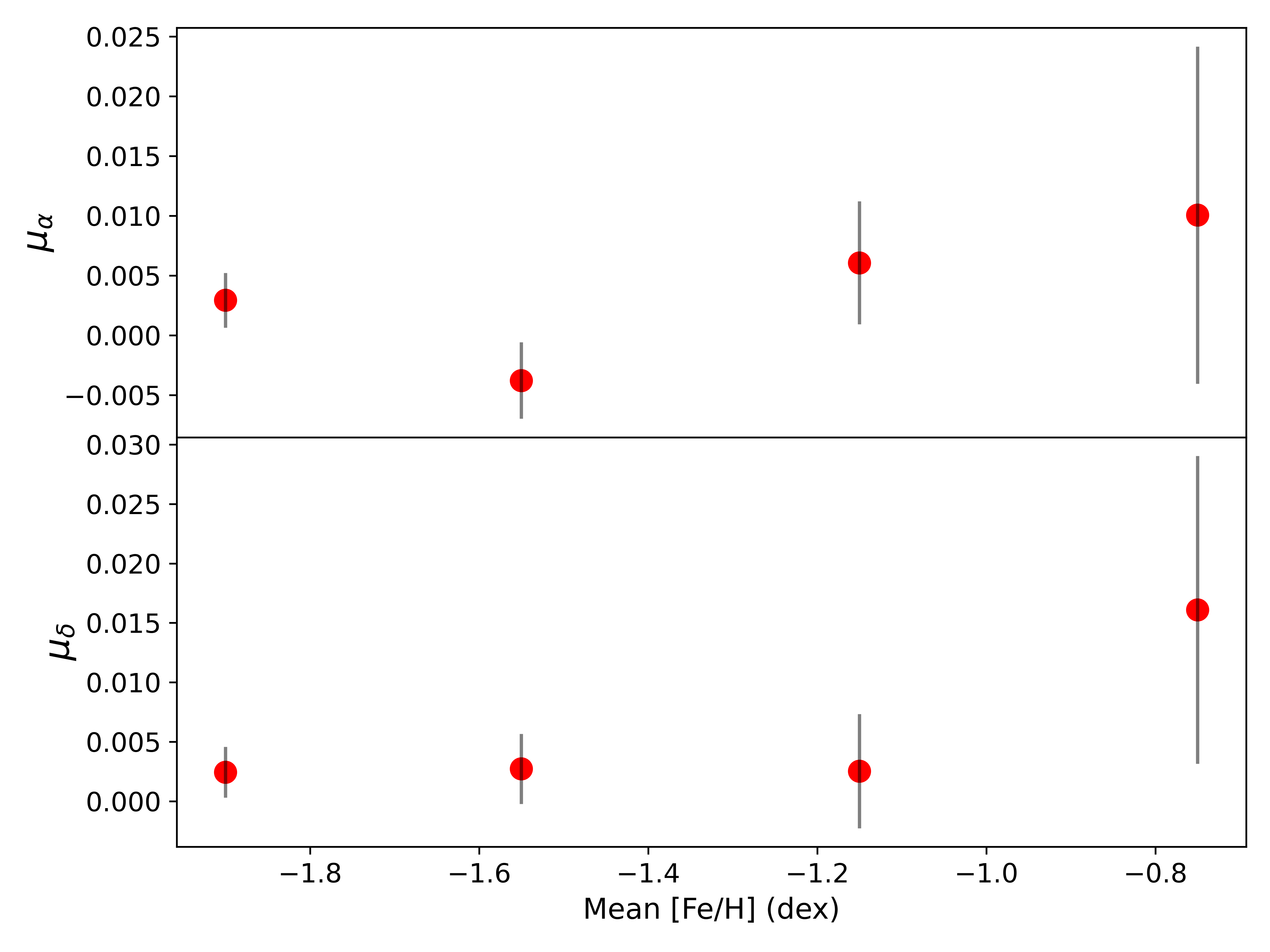}
   \caption{Mean proper motion in RA (top) and DEC (bottom) of the four sub-populations obtained using stars from oMEGACat II catalogue based on HST observations.}
   \label{fig:HST_pmr_pmt}
\end{figure}

We applied the same selection criteria on the HST astrometry as explained in \cite{omegacatII} and the same routine to obtain the cluster members. After the quality cuts, our high-quality sample had about 138\,000 stars in it, which is an order of magnitude higher than what we had in the \gaia high-quality sample. Note that the relative HST proper motions were not converted into radial and tangential components as the oMEGACat II catalogue did not include the correlation metric between the two proper motions which is essential for calculating the errors on the projections. The HST high quality sample was then divided into four sub-populations using the bounds from AG24. The number of stars in MR, MIa, MIb and MP populations are 2137, 15\,517, 40\,955 and 79\,462 stars, respectively. The percentage of stars within each population is similar to the one observed in \gaia sample. Looking at the mean proper motion values, even with a more precise dataset, we do not find a significant difference in the motions of the populations as seen in Fig. \ref{fig:HST_pmr_pmt}. We did not perform the analysis using the GMM fitting as this was supposed to be just a confirmation of the results obtained in the previous section. 

\subsection{Anisotropy}

\begin{figure*}
   \centering
   \includegraphics[width=1 \linewidth]{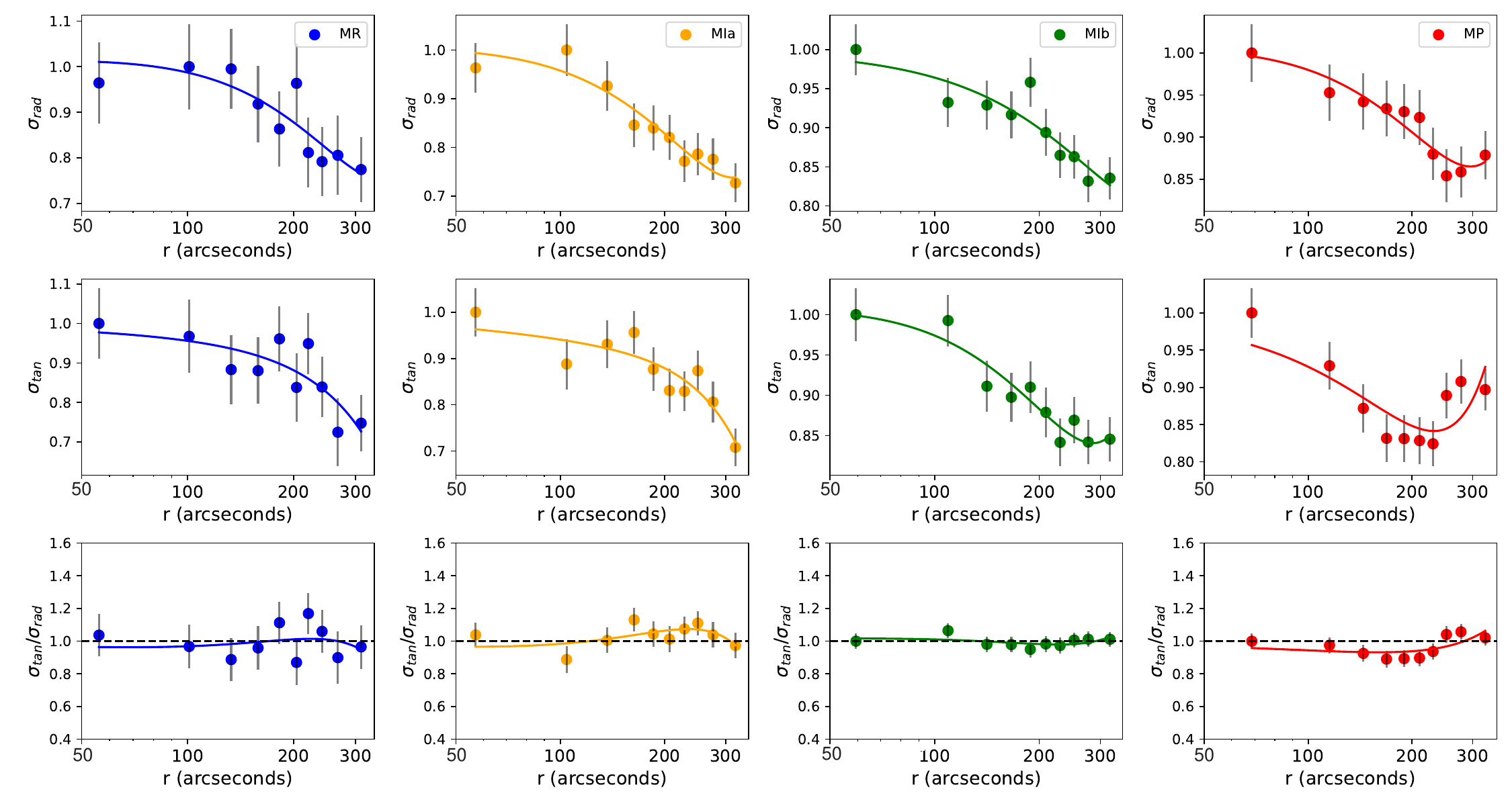}
   \caption{Velocity dispersion in radial (top row), tangential (middle row) and tangential-over-radial (bottom panel) for the four populations. The four populations, MR, MIa, MIb and MP are represented by blue, yellow, green and red circles, respectively. The solid lines are the 4th order polynomial fit to the velocity dispersion. The black line in the bottom panel represents isotropy where tangential-over-radial equals 1.}
   \label{fig:anisotropy}
\end{figure*}

\begin{figure}[h]
   \centering
   \includegraphics[width=1 \linewidth]{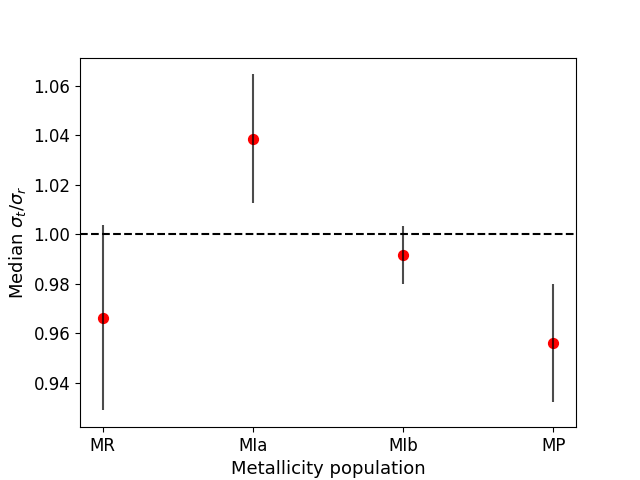}
   \caption{Median value of tangential-over-radial velocity dispersion for the four metallicity populations. The dashed black line represents isotropy.}
   \label{fig:median_iso}
\end{figure}

Anisotropy studies in GCs gain insight into their kinematics, and, when examined in terms of the different populations, they can potentially identify signatures of their formation process.\citep{tiongco2016,Tiongco2019,breen2017,breen2021,pavlik2021,pavlik2022,vesperini2021}.
We follow the method described in \cite{libralato2022} to probe the orbital anisotropies of the populations of Omega Cen. The analysis was limited to evolved stars with m$_{F625m}$ < 18.5 mag (includes sub-giant branch and red-giant branch stars). There are two reasons for this: the uncertainties on the radial and tangential proper motions are considerably larger for fainter MS stars and the offset in the metallicities of the RGB with respect to MS within the oMEGACat catalogue can lead to mixing the populations if using RGB and MS at the same time. We refined our sample by selecting stars with the highest-quality proper motion measurements: the top 20 \% of MIb and MP stars and the top 40 \% of MIa and MR stars (due to limited sample size) based on their uncertainties. The number of stars in MR, MIa, MIb and MP were 893, 1279, 5030 and 9636, respectively, with number of bins within each population set to 10, therefore each bin (except the last one) consisted of 89, 127, 503 and 963 stars, respectively. The velocity dispersion in radial and tangential directions within each bin was computed by maximising the likelihood of Eqn. 1 in \cite{libralato2022}. The posterior distributions for $\sigma_{r}$ and $\sigma_{t}$ was obtained using Markov Chain Monte Carlo (MCMC) through the {\it emcee} package \citep{emcee} with 32 walkers, 5000 steps and 500 steps as burn-in. The normalised velocity dispersion in radial, tangential and tangential-over-radial as a function of radial distance is shown in Fig. \ref{fig:anisotropy} with the median of $\sigma_{t} / \sigma_{r}$ for four populations in Fig. \ref{fig:median_iso}. All four populations show isotropy close to the center with MP showing small radial anisotropy between 150 and 200 arcseconds and MIb being isotropic throughout. Between 200 and 300 arcseconds, MIa shows tangential anisotropy but close to 300 arcseconds, all the four populations seem to show isotropy again. Looking at the median value, MR, MIb and MP show radial anisotropy, whereas MIa shows tangential anisotropy. But all four populations do not show any significant differences between each other with all the values being less than 2 $\sigma$ of each other.

\section{Discussion} \label{discussion}

Our understanding of the formation scenario of Omega Cen is still far from being complete. The cluster is often cited as an extreme sample of the complexity of the multiple population phenomenon in GCs, which has however an added degree of complexity: the MR population does not show the chemical trends and correlations which are the signature of GC stars. 
%with the origin of the MR population being particularly puzzling. In fact, unlike the other populations, it chemically does show the signatures characteristics of GC stars,  
Hypothesis for its origin have invoked accretion (either as a NSC or after being ingested into the Milky Way) or self-enrichment, with a pathway that must however have been different from that of the other populations. 
As the relaxation time of Omega Cen is relatively large for a GC, an ensamble of stars accreted by such cluster would maintain its kinematical fingerprints for a longer time and, depending on how far in the past the event took place, be detectable in the motions of such population.
%one would expect the accreted population (in this case likely the MR population) to have held on to its pre-merger kinematics. 
Our analysis of the motions of the different populations does not show significant differences for what concerns the MR population. Interestingly, the Gaia proper motions for the MIa population, %which is made up of stars that chemically behave like a GC (with anti-correlation) 
seem to have a more discrepant value -- even if still scarcely significant. This difference is not evident in the oMEGACat II data, which are however more centrally concentrated. This could possibly hint to the intriguing idea that an entire GCs was captured by a proto-Omega Cen and that the fingerprints of its motions are still detectable in the outskirts of the cluster.

The behaviour of the motions of the populations does not seem to be dependent on the metallicity bounds of the populations nor on the number of populations as the results from the 6 populations obtained through GMM fitting are also the same. 

\begin{figure}
   \centering
   \includegraphics[width=1 \linewidth]{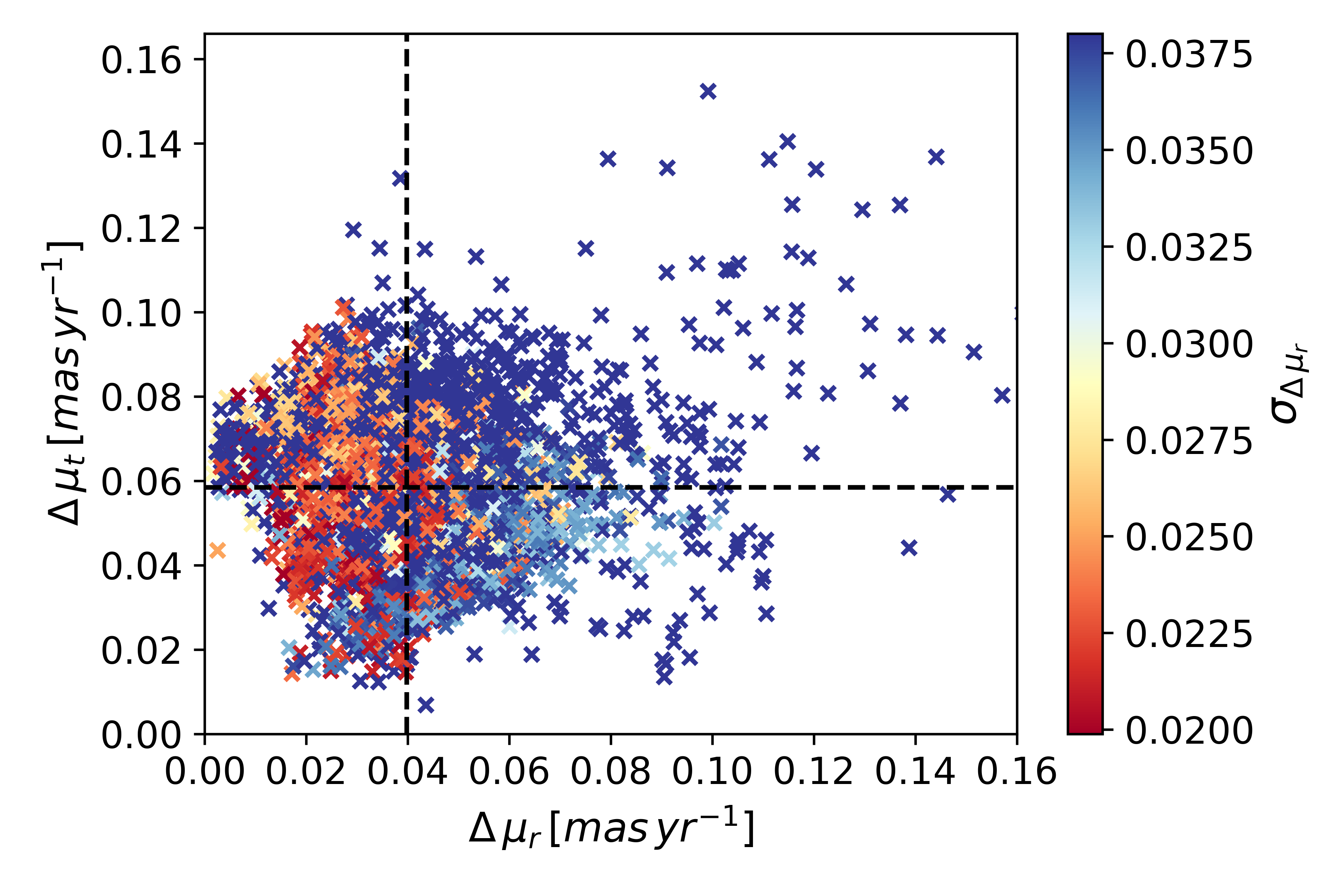}
   \caption{Largest differences in the two proper motions of the four populations. The mean of the differences are shown in black-dashed line. The colour overlay represents the uncertainties on the largest differences in the radial component of the proper motion.}
   \label{fig:motiondiff}
\end{figure}

To probe the sensitivity of the motions on the metallicity bounds, we used the method similar to Bootstrapping. We first randomly draw 50\,000 values for each of the three metallicity bounds (as there are four populations in AG24) from three pre-defined Gaussian distributions. The peaks of these distributions are adopted from AG24 with the standard deviation being kept at 0.15 dex for all the three Gaussians. Once a set of metallicity bounds is randomly selected, these bounds are used to calculate the two mean proper motions for the four populations. With the motions calculated, we measure the largest proper motion difference between the four populations and this difference is plotted in Fig. \ref{fig:motiondiff}. It is clear that the largest differences are still quite small and are usually dominated by the error on the measurement. This is a clear indication that the number of populations and bounds of the populations have very little influence on the final mean motions of the populations, 
suggesting that the motions are every uniform across the cluster.

Fig. \ref{fig:4pop} shows how the uncertainties are quite large, especially on the most metal-rich population, due to small number of stars within it. The average error on the mean motions $\mu_{r}$ and $\mu_{t}$ is about 0.05 mas. 
%This level of precision is not enough to provide us with any conclusive answer as the largest difference between the MIa and MIb has a significance of less than 2$\sigma$. 
For this reason, we also analysed the oMEGACat II catalogue, which has higher precision in comparison to the \gaia sample. The average errors on the mean motions obtained using this catalogue is about 0.005 and 0.004 mas for $\mu_{\alpha}$ and $\mu_{\delta}$, respectively (see Fig. \ref{fig:HST_pmr_pmt}). The measurements from oMEGACat II are an order of magnitude more precise than the \gaia measurements. But even with the increased precision, we do not see any significant difference between the motions of the populations with the largest difference between the population having a significance of about 1$\sigma$. Similar results were found with the anisotropy analysis, with three of the populations showing radial anisotropy but with no significant differences between each other. 

The findings on the distribution and kinematics of stellar populations in Omega Cen align with the conclusions from \cite{nitschai2024}, which observed that the metallicity populations within the cluster's half-light radius are well-mixed. This idea of well-mixed populations is further supported by the evidence of uniform rotation observed in this study. However, despite these insights, the main formation scenario of Omega Cen remains difficult to constrain due to the degeneracy between the properties of cluster stars formed through self-enrichment or through accretion, as in the latter, the stars could become kinematically integrated into the cluster. 

\section{Summary and Conclusions} \label{conclusion}

We present the kinematical analysis of stars within Omega Cen using the newly released \gaia FPR and DR3 data. The DR3\,+\,FPR sample consists about 520\,000 stars but this is reduced to about 28\,000 stars after certain quality cuts were applied to ensure only high-quality measurements were used for the analysis. The first result we present is regarding the reliability of the \gaia FPR data. We find majority of \gaia FPR stars, either located in the central region or stars that are faint, are associated with large astrometric excess noise, due to which we lose about 90$\%$ of our sample. The main result of this work is that different metallicity populations exhibit the same mean proper motions, even after using a sample larger than one used in \cite{Sanna2020} and with the inclusion of main sequence stars. This result is consistent with findings from \cite{bellini2018,libralato2018,Sanna2020}, which however used lower quality data. 
The result is not influenced by the number of populations assigned or the metallicity bounds of the populations. The third result we present is the evidence of rotation within all of the metallicity populations. They also have similar rotation amplitudes which hints towards uniform rotation among all the stars. 

Given the large uncertainties on the proper motions within the \gaia sample, we also analysed the HST sample from \cite{Haberle2024}. This new sample not only contains an order of magnitude more stars in each of the metallicity populations but also is significantly more precise than the \gaia sample. But even with all of this, we do not see a significant difference in the mean proper motions. Therefore, from previous literature and results from this analysis, the possible conclusion we can obtain regarding the formation scenario is that the cluster is either formed through self-enrichment or though accretion where the accreted population has completely integrated into the cluster. %Hopefully, the degeneracy between these two formation scenarios can be resolved with 
Future \gaia releases i.e. \gaia DR4 is expected to yield higher quality measurements even in dense fields such as in the interior of the cluster. Therefore, they might be able to provide further insight on the issue. Data from LSST will also in the future help settle this debate, given their expected high-quality proper motions for stars as faint as 27 mag in V-band.

\begin{acknowledgements}
     This work has made use of data from the European Space Agency (ESA) mission
{\it Gaia} (\url{https://www.cosmos.esa.int/gaia}), processed by the {\it Gaia}
Data Processing and Analysis Consortium (DPAC,
\url{https://www.cosmos.esa.int/web/gaia/dpac/consortium}). Funding for the DPAC
has been provided by national institutions, in particular the institutions
participating in the {\it Gaia} Multilateral Agreement. This job has made use of the Python package GaiaXPy, developed and maintained by members of the Gaia Data Processing and Analysis Consortium (DPAC), and in particular, Coordination Unit 5 (CU5), and the Data Processing Centre located at the Institute of Astronomy, Cambridge, UK (DPCI). PBK acknowledges support from the Japan Society for the Promotion of Science under the programme Postdoctoral Fellowships for Research in Japan (Standard). This work was supported by JSPS KAKENHI Grant Number JP23KF0290. 
\end{acknowledgements}

\addcontentsline{toc}{chapter}{Bibliography}
\bibliographystyle{aa}
\bibliography{reference.bib}

%\begin{appendix}
%\section{Astrometric excess noise of \gaia FPR stars}

%\begin{figure*}
 %  \centering
 %%  \caption{Astrometric excess noise of stars with respect to RA (\textit{right}) and DEC (\textit{left}). The red dashed line represents the maximum threshold for selection of the star into the high-quality sample. A peak at the center, indicating majority of centrally located stars have large astrometric excess noise.}
 %  \label{fig:astrometric}
%\end{figure*}

%\end{appendix}

\end{document}